\documentclass[sigconf, nonacm]{acmart}

\usepackage[linesnumbered,ruled,vlined]{algorithm2e}

\usepackage{url}
\usepackage{verbatim}
\usepackage{graphicx}
\usepackage{multirow}
\usepackage{balance}
\usepackage{lscape}
\usepackage{float}
\usepackage{bm}
\usepackage{xcolor}
\usepackage{paralist}
\usepackage{booktabs}
\usepackage{xspace}
\usepackage{listings}
\usepackage{lipsum}
\usepackage{subcaption}
\usepackage{xcolor}
\usepackage{xspace}
\usepackage{subcaption}
\usepackage{caption}
\usepackage{amsmath}
\usepackage{fontawesome}

\usepackage{dsfont}

\definecolor{cyanpro}{rgb}{0.0, 0.72, 0.92}
\definecolor{darkterracotta}{rgb}{0.8, 0.31, 0.36}
\definecolor{skyblue}{rgb}{0.53, 0.81, 0.92}
\definecolor{trueblue}{rgb}{0.0, 0.45, 0.81}
\definecolor{blue2}{HTML}{008AFF}
\definecolor{red3}{HTML}{CA402B}

\newcommand{\name}{\ensuremath{\mathsf{LakeVisage}}\xspace}

\newcommand{\seedb}{\ensuremath{\mathsf{SeeDB}}\xspace}
\newcommand{\tusd}{\ensuremath{\mathsf{TUS}}\xspace}
\newcommand{\lakeb}{\ensuremath{\mathsf{Lake Bench}}\xspace}
\newcommand{\santos}{\ensuremath{\mathsf{SANTOS}}\xspace}
\newcommand{\nom}{\ensuremath{\mathsf{No Merge}}\xspace}
\newcommand{\overlap}{\ensuremath{\mathsf{Overlap}}\xspace}
\newcommand{\stats}{\ensuremath{\mathsf{Stats}}\xspace}
\newcommand{\prune}{\ensuremath{\mathsf{Prune}}\xspace}
\newcommand{\company}{\ensuremath{\mathsf{Company-X}}\xspace}

\newcommand{\stitle}[1]{\vspace{0.2em}\noindent\textbf{#1}}

\newcommand{\code}[1]{\texttt{\small #1}}
\newcommand{\response}[1]{``\emph{#1}''}

\newcommand{\bigA}{\mathcal{A}}
\newcommand{\bigB}{\mathcal{B}}
\newcommand{\bigD}{\mathcal{D}}
\newcommand{\bigF}{\mathcal{F}}
\newcommand{\bigH}{\mathcal{H}}
\newcommand{\bigM}{\mathcal{M}}
\newcommand{\bigO}{\mathcal{O}}
\newcommand{\bigP}{\mathcal{P}}
\newcommand{\bigS}{\mathcal{S}}
\newcommand{\bigT}{\mathcal{T}}
\newcommand{\bigY}{\mathcal{Y}}

\newcommand\vldbdoi{XX.XX/XXX.XX}
\newcommand\vldbpages{XXX-XXX}
\newcommand\vldbvolume{14}
\newcommand\vldbissue{1}
\newcommand\vldbyear{2025}
\newcommand\vldbauthors{\authors}
\newcommand\vldbtitle{\shorttitle} 
\newcommand\vldbavailabilityurl{https://anonymous.4open.science/r/datalake-vis-3A73/}
\newcommand\vldbpagestyle{plain} 

\AtBeginDocument{%
	}
		
\begin{document}

\title{\name: Towards Scalable, Flexible and Interactive Visualization Recommendation for\\ Data Discovery over Data Lakes}

\fancyhead{}
\settopmatter{printacmref=false, printfolios=false,printccs=false
}

\author{Yihao Hu}
\affiliation{%
	\institution{Duke University}
	\country{United States}
}
\email{yihao.hu@duke.edu}
\author{Jin Wang}
\affiliation{%
	\institution{Megagon Labs}
	\country{United States}
}
\email{jin@megagon.ai}
\author{Sajjadur Rahman}
\affiliation{%
	\institution{Megagon Labs}
	\country{United States}
}
\email{sajjadur@megagon.ai}

\begin{abstract}
	Data discovery from data lakes is an essential application in modern data science.
	While many previous studies focused on improving the efficiency and effectiveness of data discovery, little attention has been paid to the usability of such applications.
    In particular, exploring data discovery results can be cumbersome due to the cognitive load involved in understanding raw tabular results and identifying insights to draw conclusions.
    To address this challenge, we introduce a new problem --- visualization recommendation for data discovery over data lakes --- which aims at automatically identifying visualizations that highlight relevant or desired trends in the results returned by data discovery engines. 
	We propose \name, an end-to-end framework as the first solution to this problem.
	Given a data lake, a data discovery engine, and a user-specified query table, \name intelligently explores the space of visualizations and recommends the most useful and ``interesting'' visualization plans.
	To this end, we developed (i) approaches to smartly construct the candidate visualization plans from the results of the data discovery engine and (ii) effective pruning strategies to filter out less interesting plans so as to accelerate the visual analysis.
	Experimental results on real data lakes show that our proposed techniques can lead to an order of magnitude speedup in visualization recommendation. 
	We also conduct a comprehensive user study to demonstrate that \name offers convenience to users in real data analysis applications by enabling them seamlessly get started with the tasks and performing explorations flexibly.
\end{abstract}
\maketitle

\pagestyle{\vldbpagestyle}
\begingroup\small\noindent\raggedright\textbf{PVLDB Reference Format:}\\
\vldbauthors. \vldbtitle. PVLDB, \vldbvolume(\vldbissue): \vldbpages, \vldbyear.\\
\href{https://doi.org/\vldbdoi}{doi:\vldbdoi}
\endgroup
\begingroup
\renewcommand\thefootnote{}\footnote{\noindent
	This work is licensed under the Creative Commons BY-NC-ND 4.0 International License. Visit \url{https://creativecommons.org/licenses/by-nc-nd/4.0/} to view a copy of this license. For any use beyond those covered by this license, obtain permission by emailing \href{mailto:info@vldb.org}{info@vldb.org}. Copyright is held by the owner/author(s). Publication rights licensed to the VLDB Endowment. \\
	\raggedright Proceedings of the VLDB Endowment, Vol. \vldbvolume, No. \vldbissue\ %
	ISSN 2150-8097. \\
	\href{https://doi.org/\vldbdoi}{doi:\vldbdoi} \\
}\addtocounter{footnote}{-1}\endgroup

\ifdefempty{\vldbavailabilityurl}{}{
	\vspace{.3cm}
	\begingroup\small\noindent\raggedright\textbf{PVLDB Artifact Availability:}\\
	The source code, data, and/or other artifacts have been made available at \url{\vldbavailabilityurl}.
	\endgroup
}

\section{Introduction}\label{sec-intro}

In the past decades, there has been a drastic growth in the number of open and shared datasets from governments, academic institutions, and enterprises.
These massive collections of datasets, known as \emph{data lakes}, open up new opportunities for innovation, economic
growth, and social benefits~\cite{DBLP:conf/sigmod/Fan00M23,DBLP:journals/debu/MillerNZCPA18}.
\emph{Data discovery}, which aims at helping users find and access specific data they want, is an essential operation for integrating and analyzing the datasets from data lakes that are useful for various downstream tasks.
Data discovery has become an important topic in the data management community with a specific focus on topics such as data exploration~\cite{DBLP:conf/icde/FernandezAKYMS18,DBLP:conf/icde/SantosBMF22,DBLP:conf/icde/GalhotraGF23}, table union search~\cite{DBLP:journals/pvldb/NargesianZPM18,DBLP:journals/pacmmod/KhatiwadaFSCGMR23,DBLP:journals/pvldb/FanWLZM23}, joinable table discovery~\cite{DBLP:conf/sigmod/ZhuDNM19,DBLP:journals/pvldb/EsmailoghliQA22,DBLP:journals/pvldb/Dong0NEO23}, and domain discovery~\cite{DBLP:journals/pvldb/ZhuNPM16,DBLP:journals/pvldb/OtaMFS20}.

\begin{figure}[h!t]
	\centering
	\includegraphics[width=0.48\textwidth]{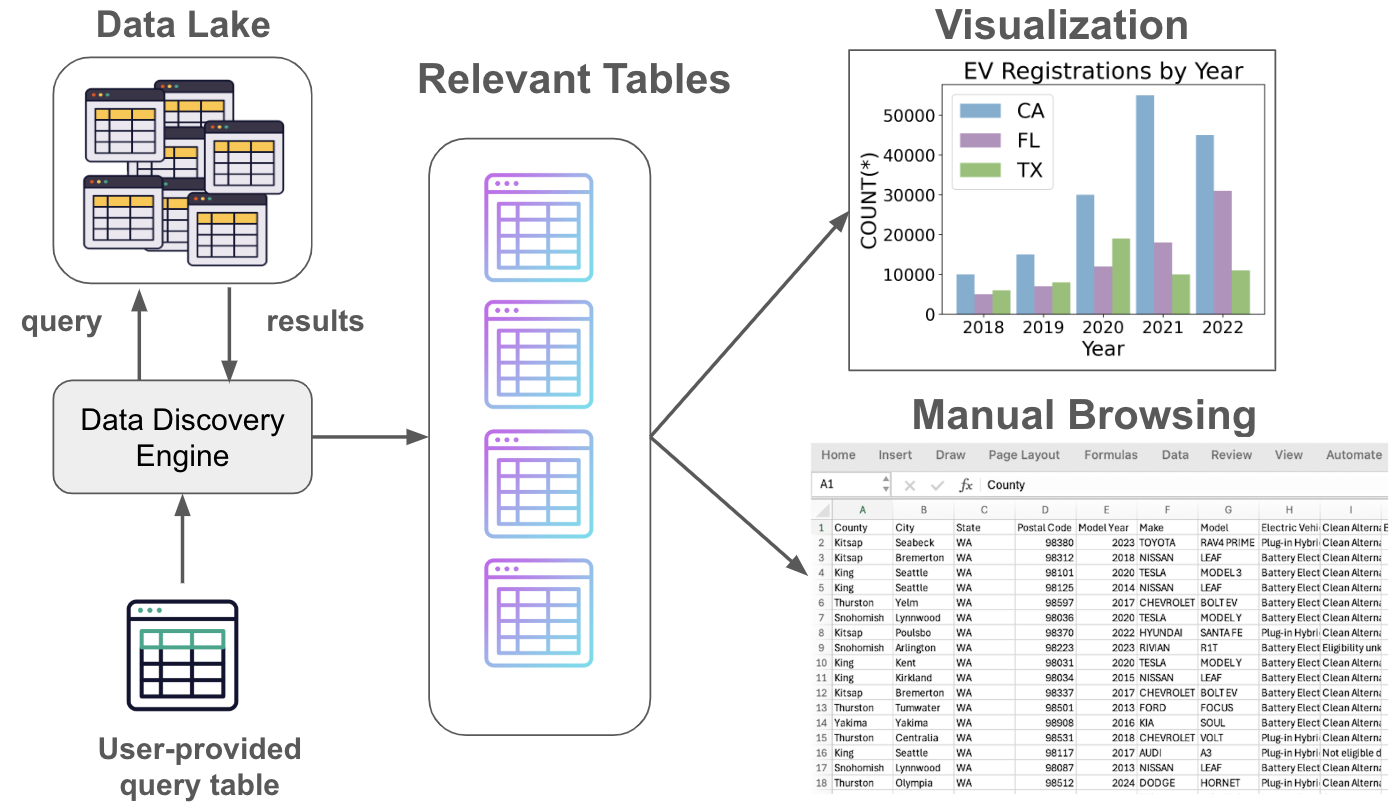}
	\caption{Example:  Analyzing Data Discovery Results.}
	\label{fig:motivation}
\end{figure}

While many previous studies focus on improving the efficiency and effectiveness of data discovery, little attention is paid to its usability.
Since the objective of data discovery is to provide richer data to boost the data analysis tasks, it is essential to make the discovery results easily accessible and usable.
In practical scenarios, the data discovery outcomes are always collections of tables with heterogeneous schema, wherein a subset of columns or rows of different tables are related~\cite{DBLP:conf/www/BrickleyBN19,DBLP:journals/pvldb/LehmbergB17, DBLP:conf/sigmod/YakoutGCC12,DBLP:conf/sigmod/SarmaFGHLWXY12}. 
These relations often represent semantic relatedness obtained via approximations.
Existing studies on data discovery put the onus on the user to analyze and make sense of such results as simply the raw tables are returned for a given query.
Therefore, exploring a collection of tables returned as data discovery results can be even more challenging.
Although there have been some efforts in easing the exploration of data lakes~\cite{DBLP:conf/icde/FernandezAKYMS18,DBLP:conf/sigmod/NargesianPZBM20,DBLP:journals/pvldb/OuelletteSNBZPM21}, they target at improving the usability of browsing the whole data lake instead of the data discovery results.
Therefore, they cannot be utilized to resolve above challenges. 
To address this issue, a natural approach is to provide visualizations, which is often the first step in data analysis~\cite{DBLP:journals/tvcg/WongsuphasawatM16,DBLP:journals/pvldb/VartakRMPP15,DBLP:journals/tvcg/ElmqvistF10,DBLP:journals/pvldb/LeeTABCKMSYHP21}, as affordances to interactively explore data discovery results.
With the help of visualization, it would be much easier for users to analyze the data discovery results.
A typical example is shown in Figure~\ref{fig:motivation}:

\begin{example}
Suppose Aluko, a data scientist, has a dataset about national electric vehicle registration and wants to gain more insights to serve for potential business decision making. 
To this end, she first tries to finds a list of tables that are relevant to that dataset, which is considered as the query table, via a data discovery engine and then analyzes the collection of tables to derive insights.

Suppose the schema of query table includes columns like location, date, model, make, and color, among others, and the result tables have similar but not exactly the same schema. If she is interested in finding the year with the largest registration increase for different states, e.g., CA, here are some necessary steps she has to take:
\begin{enumerate}
    \item Perform data cleaning on all table schemas and table contents to ensure consistent data quality across all tables.
    \item Collect metadata (types, value range, etc.) for columns in all tables to properly align them for further analysis.
    \item Write programs to perform the designated analysis task.
\end{enumerate}
When relying on manual browsing, each of the above steps can potentially require significant coding or locating efforts, especially when the data size is huge. On the contrary, \name takes all the tables returned by the search engine, strategically resolves column alignments, and searches for and presents relevant and potentially interesting visualizations, freeing users from any tedious work and making the data exploration easier.
\end{example}

Many previous works provide scalable and interactive visualization recommendations for structured queries over relational data~\cite{DBLP:journals/pvldb/KahngNSC16,DBLP:journals/pvldb/SiddiquiKLKP16,DBLP:journals/pvldb/KimBPIMR15,DBLP:journals/pvldb/VartakRMPP15,DBLP:journals/pvldb/LeeTABCKMSYHP21}.
Given a structured query such as SQL, they can automatically identify the most interesting visualizations for analytical tasks. 
However, it is non-trivial to extend these techniques to the scenario of data discovery tasks over data lakes due to the following reasons:
Firstly, since data discovery tasks aim at finding related tables from the data lake, it is essential to reflect the relationship between query and result tables in the visualization.
Thus a new problem definition considering such requirements is needed.
In addition,  visualizations of multiple related columns among the data discovery results can introduce \emph{perceptual scalability} challenges for the users, which exists when analyzing even a single table, i.e., viewing and exploring the information in such tables put the cognitive burden on users~\cite{DBLP:journals/tvcg/YostN06,DBLP:journals/tvcg/ElmqvistF10,DBLP:journals/pvldb/RahmanBLZSKP21}. 
Unlike previous work for relational databases, the visualization needs to present information from multiple tables in data discovery scenarios.
Thus it is essential to resolve the challenge of perceptual scalability in this process.
Moreover, while the results of SQL queries are always a set of tuples, those of data discovery tasks are collections of tables.
As a result, the search space of the visualization recommendation for data lakes would be much larger than that of relational databases.
To ensure the user experience, it is essential to reduce the latency of the recommendation process.

In this paper, we study how visualization recommendation can help ease the exploration of the data discovery results from data lakes. 
To this end, we propose an end-to-end framework (\name\footnote{LakeVisage is a portmanteau of the words lake (from data lakes) and envisage, thereby representing the goal of helping users comprehend data discovery results visually.}) that recommends visualizations as affordances to assist users in seamlessly exploring the results returned by a data discovery engine.
Since there is no previous study on this problem, we first come up with a formal definition for it to illustrate how to build visualization over results across multiple tables.
To illustrate the relatedness between tables in the results and the given query, we need to allocate result tables of data discovery into a certain number of \emph{series} in the visualization, which could also help improve perceptual scalability.
We show that this problem is intractable and provide a data-driven solution to efficiently generate high-quality results based on the data distribution of involved columns.
We also develop effective pruning techniques to reduce the cost of handling unpromising visualization plans and accelerate the overall query processing time.
These techniques could result in up to 30X of performance gain in total. 
Our contributions in this paper are summarized as follows:
\begin{compactitem}
	\item We study the new research problem of visualization recommendation for data discovery over data lakes and come up with the first formal definition for it.
	\item We design and implement \name, an end-to-end framework, as the solution to this problem. Specifically, we develop novel techniques to support constructing visualizations over multiple result tables and propose a suite of pruning techniques to reduce the execution time.
	\item We conduct comprehensive experiments over three public datasets of data discovery. The results show that our proposed techniques could reduce the overall execution time by an order of magnitude.
	\item We present the results of a systematic user study aimed at evaluating the usability and effectiveness of \name compared to Jupyter Notebooks, a literate programming tool widely used by practitioners for exploring data discovery results. The outcome of user study further justifies that the visualization generated by our proposed framework could significantly ease the data analysis of data lake-related tasks.
\end{compactitem}

 The rest of this paper is organized as follows: 
 Section~\ref{sec-prelim} introduces the necessary background knowledge and formally defines the problem.
 Section~\ref{sec-method} presents our proposed framework with technical details.
 Section~\ref{sec-disc} discusses some essential topics regarding our proposed framework's flexibility and potential extensions.
 Section~\ref{sec-exp} illustrates the experimental results over real benchmarking datasets.
 Section~\ref{sec-user} introduces the design and results of the user study.
 Section~\ref{sec-related} surveys the related work, and Section~\ref{sec-conc} contains concluding remarks.

\section{Preliminary}\label{sec-prelim}


\subsection{Data Discovery}\label{subsec-disc}

The core task of data discovery is finding related tables~\cite{DBLP:conf/sigmod/SarmaFGHLWXY12} from data lakes for a given query table. 
Suppose the data lake is a collection of tables $\bigT$, and each table $T \in \bigT$ consists of several columns. 
To decide whether two tables are related, we need to compute the relevant scores between all pairs of columns from them and then aggregate them to decide the table relatedness.
There are many definitions of table relatedness, and in this paper we will focus on the problem of Table Union Search~\cite{DBLP:journals/pvldb/NargesianZPM18} as an example to illustrate our proposed techniques, which could be easily extended to support other data discovery tasks.
While many previous studies~\cite{DBLP:journals/pvldb/NargesianZPM18,DBLP:conf/icde/BogatuFP020,DBLP:journals/pacmmod/KhatiwadaFSCGMR23,DBLP:journals/pvldb/FanWLZM23,DBLP:conf/icde/FanSM24} have proposed different methods to decide the unionability between tables, we can just use the notation $U$ to denote it. 
And thus a general definition of the Table Union Search problem could be illustrated as Definition~\ref{def-tus}.

\begin{definition}[Table Union Search] \label{def-tus}
	Given a collection of data lake tables $\bigT$  and a query table $S$, top-k table union search aims at finding a subset $\bigS \subseteq \bigT$ where $|\bigS| = k$ and $\forall T \in \bigS$ and $T' \in \bigT- \bigS $, we have $U(S, T) \geq U(S, T')$.
\end{definition}

Note that our proposed framework could apply to any result set irrespective of the underlying Table Union Search method and does not need to be aware of the corresponding unionability scoring metrics.
Besides, since other data discovery tasks can also be formalized into the similar data model, we will use the term ``data discovery'' when introducing the proposed techniques in the rest of this paper.
In addition, since our proposed techniques are also independent from the visualization format, we will use bar charts as an example when introducing our framework.

\subsection{Visualization Recommendation}\label{subsec-vis}

We now introduce the essential terminologies of visualization recommendation. 
A representative work in this field is \seedb~\cite{DBLP:journals/pvldb/VartakRMPP15}, which focuses on visualization recommendations for a relational database $D$ with a snowflake schema.
In visualization recommendation literatures, the setting is assumed to be database $D$ with a snowflake schema.
Within such a setting, there are three key aspects for visualization generation: \emph{dimension attributes} $\bigA$ are the attributes to group-by in a visualization; \emph{measure attributes} $\bigM$ are the attributes to perform aggregate on in the visualizations; and aggregate functions $\bigF$ (such as SUM, COUNT, and AVG) over measure attributes.
Therefore, a \emph{visualization plan} $P$ is a function represented by a triple $\langle A, M, F \rangle$, where $A \in \bigA$, $M \in \bigM$, $F \in \bigF$. 
Actually, $P$ could be regarded as a two-column table $\langle A, F(M) \rangle$, which can be displayed via standard visualization mechanisms, such as bar charts or trend lines. 
The visualization recommendation approaches typically focus on such visualization plans due to their reported popularity among users of visualization tools~\cite{DBLP:journals/pvldb/VartakRMPP15,DBLP:journals/pvldb/MortonBGM14}. 
Given a user query $Q$ that aims to explore a subset of table $D$ and results in a visualization $P_Q$, the goal of visualization recommendation is to identify other visualizations that are interesting with respect to $P_Q$ on a predefined utility function.
Examples of interestingness include larger deviation from the distribution underlying $P_Q$, which could be measured via utility functions such as Earth Mover's Distance (EMD), Euclidean Distance, Kullback-Leibler Divergence (K-L divergence),and Jenson-Shannon Distance~\cite{DBLP:journals/pvldb/VartakRMPP15,DBLP:conf/iui/LeeDHEP19}. 

\subsection{Problem Definition}\label{subsec-problem}

Next, we will explain the formal definition of the problem visualization recommendation for data discovery using the above-mentioned terminologies.
Let $Q$ denote the query table and $\bigS = \{S_1, S_2,...,S_k \}$ denote the set of $k$ result tables of the Table Union Search problem, respectively.
Suppose $Q$ has $u$ columns where the $i^{th}$ column is denoted as $q_i$, and the $x^{th}$ column of the $y^{th}$ result table $S_y $ is denoted as $s_{xy}$.
Based on the definition of data discovery, each column $q \in Q$ is aligned with a set of columns in the result tables denoted as $C(q)$.
For each $C(q)$, we have $|C(q)| \leq k$ since each column $q \in Q$ will align with at most one column in each result table $S \in \bigS$.

\begin{figure}
	\centering
	\includegraphics[width=0.45\textwidth]{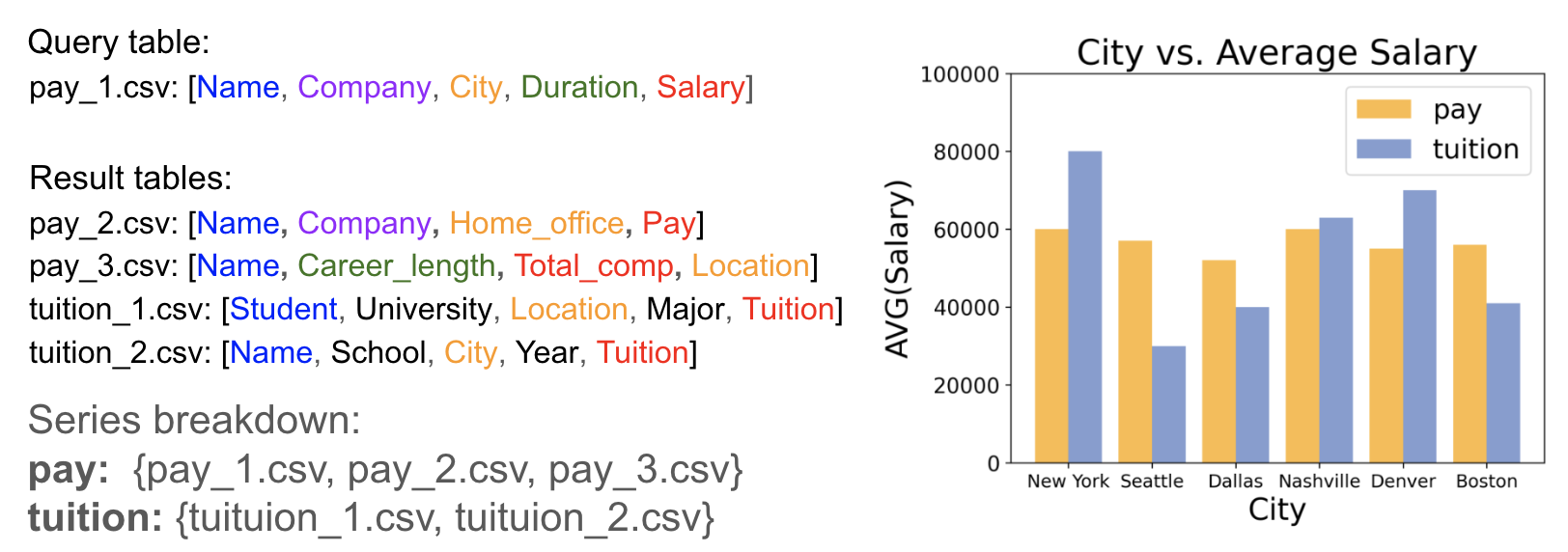}
	\caption{Illustration of Terminologies for Visualization}
	\label{fig:ex2}
\end{figure}

Regarding the definition of visualization plans, we follow the idea of related work introduced earlier by denoting a visualization plan using the triplet $P = \langle A, M, F \rangle$.
While the visualization plans for relational databases in previous studies introduced above are two-column tables $\langle A, F(M) \rangle$, in the scenario of data discovery, the visualization needs to display results from multiple tables, $k$ to be precise.
Besides,  since data discovery tasks aim at identifying tables relevant to the given query, it is essential to illustrate such relatedness in visualization plans.
To satisfy this requirement, we propose the new concept of \emph{series} that can be formally described as following: 
given a query column $q$ and the set of its aligned columns $C(q)$, a \emph{series} $\gamma$ is the ordered vertical concatenation of $w$ aligned columns that are related with each other in $C(q)$ where $w \in [1, |C(q)|]$.
We denote the set of series corresponding to $C(q)$ is as $\Gamma(q)$, where $1 < |\Gamma(q)| \leq |C(q)|$. 
And the set of all series for the query table $Q$ is denoted as $\Gamma_Q = \cup_{q \in Q} \Gamma(q)$.
In the visualization plan for data discovery, the bars under a value of dimension attribute $A$ should correspond to a series $\gamma \in \Gamma_Q$, while the values of $A$ should come from the union of all columns in $C(q)$ and $q \in Q$.
In this way, we could also improve the perceptual scalability since the number of series is no larger than that of aligned columns in result tables.
The definitions of the measurement attribute $M$ and aggregate $F$ are similar to previous work, and $F(M)$ returns a single real number.
Here, we assume that $F$ includes only one aggregation operation on one column for ease of presentation. 

\begin{example}
	We show a running example to illustrate above terminologies.
	In the upper left part of Figure~\ref{fig:ex2}, it shows an example of a query table and its result tables, where the aligned columns are in the same color.
	For example, for the query column $q$ of \emph{Salary}, the set of aligned columns $C(q)$ is $\{Pay, Total\_comp, Tuition, Tuition\}$.
	The right part of Figure~\ref{fig:ex2} presents a valid visualization plan of the query and result tables.
	Here the dimension attribute $A$ is \emph{City} and there are 6 values.
	The set of series for above $C(q)$ is $\Gamma = \{pay, tuition\}$ as shown in bottom left of the figure, where \emph{pay} is the union of column \emph{Salary} in table \emph{pay\_1.csv}, column \emph{Pay} in table \emph{pay\_2.csv} and column \emph{Total\_comp} in table \emph{pay\_3.csv}; \emph{tuition} is the union of column \emph{Tuition} in table \emph{tuition\_1.csv} and table \emph{tuition\_2.csv}.
	The measure attribute $M$ is \emph{Salary} and the aggregate $F$ is AVG.
\end{example}

Following the practice of previous studies~\cite{DBLP:conf/iui/LeeDHEP19,DBLP:journals/pvldb/VartakRMPP15}, we also use Earth Mover's Distance (EMD) as the utility function $\bigD$ to evaluate whether a visualization plan is interesting. 
Note that in keeping with classical visualization recommendation literature, the focus of this work is to provide a framework that facilitates integration of any suitable metric rather than find the best utility metric.
Suppose there are $v$ series in $\Gamma$, the \emph{utility score} of a visualization plan $P$ can be calculated with Equation~\eqref{eq-emd}:
\begin{equation}\label{eq-emd}
	\bigD(P) = \frac{2}{v*(v-1)} \sum\limits_{i,j \in [1, v], i \neq j} EMD(\gamma_i, \gamma_j) 
\end{equation}
here we use $\gamma_i$ interchangeably to denote the associated numerical vector of a series; the vector dimension equals the number of unique values in $A$, and the value in each dimension is the result of $F(M)$ for each value in $A$.
Suppose we read the values of two series for all locations from Figure~\ref{fig:ex2} and formulate the vectors $[60009.24, 56940.46, 52323.85, 60146.35, 55112.33, 56197.89]$ and $[79794.62, 30446.72, 39829.39, 62852.72, 69914.3, 40512.97]$, respectively. 
Then the utility score of this plan could be calculated by first obtaining the distribution of each array via normalization.
Given these distributions, we directly calculate the EMD to be $0.16$.

Finally we have our formal problem definition of visualization recommendation for data discovery as Definition~\ref{def-main}:

\begin{definition}\label{def-main}
	Given a query table $Q$ and its set of result tables $\bigS$, the visualization recommendation for data discovery problems aims at finding top-$n$ most interesting visualization plans $\bigP$ where $\forall P \in \bigP, P' \notin \bigP$ we have $\bigD(P) \geq \bigD(P')$
\end{definition}

\section{Methodology}\label{sec-method}

\subsection{System Overview}\label{subsec-sys}

\begin{figure}[h]
	\centering
	\hspace{-1.15em}\includegraphics[width=0.5\textwidth]{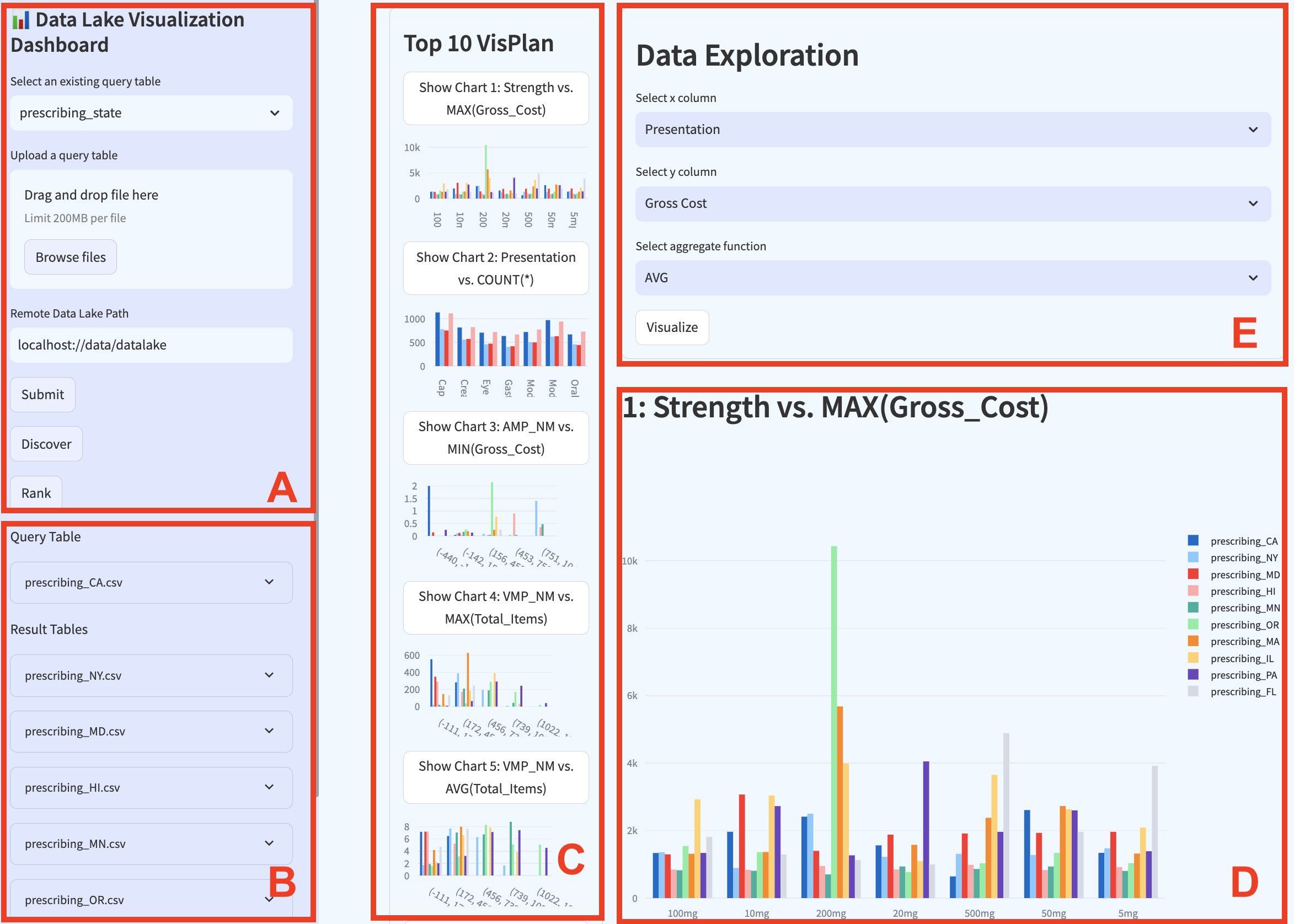}
	\caption{The Front-end of \name.}
	\label{fig:ui}
\end{figure}

We first introduce the front-end of \name as shown in Figure~\ref{fig:ui}.
It provides an interactive user interface that allows users to specify the visualizations and returns the top-ranked visualization plans recommended by our proposed algorithms.
The web-based front end consists of five components: (A) a query builder where the user specifies the query table and path to the data lake storage; (B) a schema viewer which displays the schema information of query and result tables; (C) a recommendations panel which displays the recommended visualization plans, (D) a detailed view which highlights the plan a user selected from the recommendations; and (E) a plan builder for users to specify customized visualization plans.


\begin{figure}[h]
	\centering
\vspace{-10pt}	\hspace{-1.15em}\includegraphics[width=0.5\textwidth]{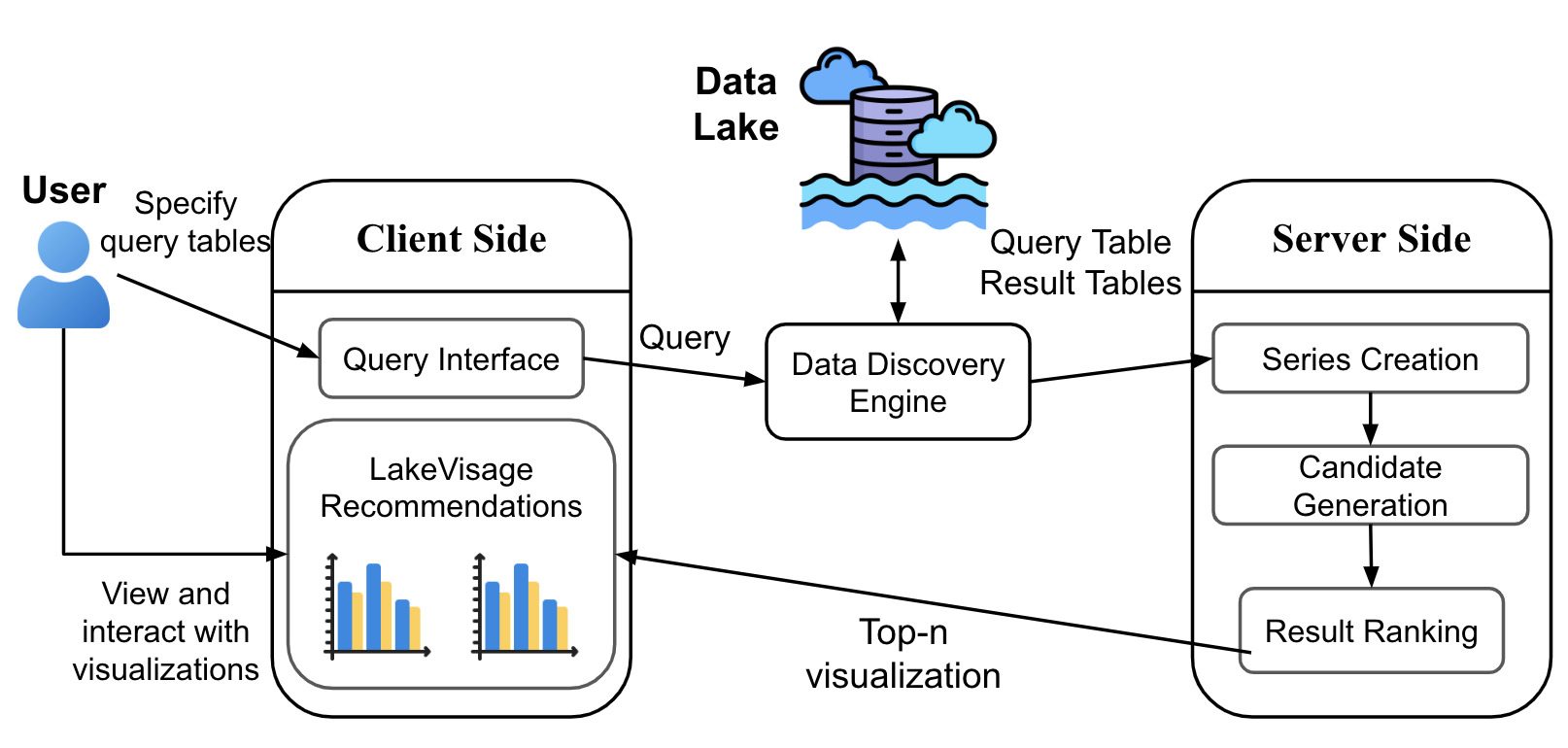}
	\caption{The Overall Architecture.}
	\label{fig:framework}
\end{figure}

\name supports an end-to-end pipeline that could be built on top of any data discovery engine.
The overall workflow is shown in Figure~\ref{fig:framework}.
Given the input table, the data discovery framework first retrieves the top-$k$ relevant result tables from the data lake.
Then there are three major steps for the visualization recommendation process: \emph{Series Creation}, \emph{Candidate Generation} and \emph{Results Ranking}.
The Series Creation step aims at creating the series to serve as the bars denoting the $F(M)$ values in the visualization chart for each dimension attribute values;
The Candidate Generation Step tries to identify promising visualization plans from the large search space to reduce the overhead of computing utility scores.
The Results Ranking step verifies the utility score for all candidate plans and returns the top-$n$ ones as the recommendation results.

One important issue to be resolved for the data lake scenarios is to handle the heterogeneous data formats.
\name supports three data types: categorical, numerical, and textual.
Based on the definition of visualization plans, we only consider categorical, numerical columns as the candidates for measuring 
attributes ($M$) in a plan.
When a column appears as a dimension attribute ($A$), we transform its entries, i.e., cells in the table column, to categorical values when they are not.
We discretize the column values into several bins for numerical columns and map each entry to the respective bins.
Each bin corresponds to a value of the dimension attribute.
For textual columns, we transform each entry into a low-dimensional embedding vector.
Then we perform a clustering over the embedding of all cells and treat each cluster as a dimension attribute value.
Similar discretization approaches have been employed by tools for interactive analytics on textual data~\cite{DBLP:conf/cidr/RahmanGD21,DBLP:conf/chi/ZhangEELDT20}.
Note that the discretization approaches may vary depending on the data (e.g., reviews, QA) and downstream task (e.g., grouping/clustering, sentiment analysis, and  natural language inference, among others.) 
While we currently integrated only clustering-based approaches, \name can be extended to support a wide variety of discretization techniques via user-defined functions.

\subsection{Series Creation for Visualization}\label{subsec-series}

In order to satisfy the requirements of showing the relatedness between tables in data discovery results, we need to construct high-quality series collection $\Gamma(q)$ for each $q \in Q$ from its aligned columns $C(q)$ introduced before to serve as the bars for each attribute dimension value in the visualization chart.
The general guidelines for series creation include: (i) the columns belonging to the same series should be closely related; (ii) the overall utility score of the series over different attribute dimension values should be large to make more contributions to the plan; and (iii) there should not be too many series to ensure the perceptual scalability.

Among the above items, the second and third ones could be satisfied during the Result Ranking step.
And the series creation process needs to focus on the first one.
In the simplest scenario, each column $s_{xy} \in C(q)$ is regarded as a series.
However, such a case may lead to a loss of context regarding the relatedness of columns. Moreover, as the number of series increases, users are overloaded with information leading to perceptual scalability challenges.
To acquire high-quality series with relevant columns, we need to define a mechanism for evaluating their relatedness. 
We start from a heuristic method based on syntactic similarity as the baseline: we consider columns with similarity scores larger than a pre-defined threshold as related ones.
Specifically, we consider the criteria of columns with different data formats as follows:
\begin{compactitem}
	\item Categorical: Regard two columns as two sets and consider Jaccard similarity between them
	\item Numerical: Identify the range of two columns and consider their overlaps
	\item Textual: Regard two columns as two bags of words and consider Jaccard similarity between them
\end{compactitem}

Nevertheless, the above syntactic similarity might not be an ideal solution due to two concerns: on the one hand, the simple similarity scores failed to carry enough semantic information among multiple columns within a series.
On the other hand, traversing two columns is required to compute the similarity score. 
Thus, the time complexity is related to the size of tables the columns belong to, which is expensive considering the large size of data lake tables.
To address such issues, we introduce a data-driven approach with the help of statistical tools.
The high-level idea is to treat each column as a random variable with values for each entry drawn from a distribution.
Then we can decide whether two columns should belong to one series based on their distributions.
Given two columns~\footnote{The comparison could also happen between two series where each series is a combination of multiple columns. We will only mention columns when introducing the techniques for the ease of presentation.}, if their distributions are close enough, they will be considered as belonging to one series and thus merged into one series.
To reach this goal, we employ \textbf{Pearson’s chi-square test}~\cite{gupta2020fundamentals}, which measures whether the observed frequency distribution of a random variable is significantly different from its expected frequency distribution.
The statistics is calculated as Equation~\eqref{eq-chi}:
\begin{equation}\label{eq-chi}
	\chi^2 = \sum_{i=1}^N \frac{(O_i - E_i)^2}{E_i}  
\end{equation}
where $N$ is the number of categories, $O_i$ and $E_i$ is the observed and expected frequency of each category.

In our scenario, we can regard the two columns as the observed and expected distribution, respectively and apply such a tool.
For the column regarded as observed one, we conduct random sampling to obtain $W$ samples to serve as the observations $O_i$;
For the column regarded as the expected one, we need to estimate its statistics to decide the above value of $E_i$.
We assume that the expected column follows the exponential distribution family~\footnote{https://en.wikipedia.org/wiki/Exponential\_family}, which covers most common distributions such as Bernoulli, normal, and gamma distribution.
The statistics could then be estimated by randomly selecting $W$ samples from the column and then conducting Maximum Likelihood Estimation.
We empirically observed that setting $W$ to 500, which are less than 1\% of most tables, yielded very promising results.
After obtaining values of $O_i$ and $E_i$ with the above sampling and estimation approach, we can then calculate the $\chi^2$ test statistic and then decide whether to reject the null hypothesis, i.e., the two data distributions are similar, accordingly.
Note that the overhead of estimating $E_i$ values would be trivial.
Once we decide which columns belong to the same series, we can decide the statistics of the newly formed series based on those of the two columns.
For instance, suppose two columns $c_1$ and $c_2$ follow Gaussian distributions $N(\mu_1, \sigma_1^2)$ and $N(\mu_2, \sigma_2^2)$, respectively, then the series consists of them will follow the  Gaussian distributions $N(\mu_1+\mu_2, \sigma_1^2+\sigma_2^2)$.
Thus, the required number of estimations for a given column $q \in Q$ is no more than the cardinality of its aligned columns $|C(q)|$.


Another issue to be resolved in series creation is the large search space.
Given the column $q \in Q$, the number of potential candidates for series collection is $\bigO(2^n)$, where $n$ is the number of aligned columns in result tables ($|C(q)|$).
To reduce the computation overhead, we propose an iterative process for finding promising series collections without traversing the whole search space.
The high-level idea is that we sort all columns in $C(q)$ in ascending order of cardinality and initialize each series with one single column.
Then, we start from the smallest columns and conduct the above statistical tests on a pair of columns.
If they pass the test, they will be merged into one new series and proceed to the next round.
In each round, we will only merge one series pair.
The whole process will end if there are only two series left or no pair of series can be further merged.
This process could be efficiently implemented with data structures like Union Find.

\SetKwInOut{Parameter}{Variables}
\begin{algorithm}[!t]
	\small
	\KwIn{The Query table $Q$; The set of result tables $\bigS$}
	\KwOut{The Series Collection $\Gamma_Q$}
	Initialize $\Gamma_Q = \emptyset$\;
	\ForEach{$q \in Q$}{
		Initial $\Gamma(q) = \emptyset$\;
		Obtain $C(q)$ from result tables $\bigS$\;
		Sort $C(q)$ in ascending order of cardinality\;
		\For{$c_j \in C(q)$}{
			$\Gamma(q) \leftarrow \Gamma(q) \cup c_j$\;
		}
		Initialize cursor $j = 0$\;
		\While{$|\Gamma(q)| > 2$ and $j < |\Gamma(q)|$}{
			Conduct statistic test on $\Gamma(q)[j]$ and $\Gamma(q)[j+1]$\;
			\uIf{Test pass}{
				Merge them into a new series, update $\Gamma(q)$\;
				Reset $j = 0$\;
			} 
			\Else{
				$j += 1$\;
			}
		}
		$\Gamma_Q \leftarrow \Gamma_Q \cup \Gamma(q)$\;
	}
	\Return $\Gamma_Q$\;
\caption{Series Creation}
\label{alg:series}
\end{algorithm}	

Based on the above discussion, we propose the solution for series creation as shown in Algorithm~\ref{alg:series}.
The complexity could be analyzed as follows:
First, $\forall q \in Q$ the cardinality of $|C(q)|$ is $\bigO(k)$ based on the problem setting where $k$ is the number of result tables.
The time to sort each $C(q)$ is $\bigO(k \log k)$.
Given two columns in $C(q)$, the time to perform chi-square test is $\bigO(W)$ where $W$ is the sample size;
and the number of test is $\bigO(k)$.
Thus the time complexity for handling each $C(q)$ is $\bigO(k \log k+k*W)$, while the overall time complexity of the $u$ columns in the query table would be $\bigO(u*k*(\log k +W))$.
The overall computation overhead would be trivial since $W$ and $k$ will be relatively small numbers.

\subsection{Efficient Candidate Generation}\label{subsec-opt}

After creating the series for query and result tables, we can construct the visualization plans accordingly.
As discussed before in Section~\ref{subsec-problem}, each plan is a triplet $P = \langle A, M, F\rangle$.
Here the potential candidate of dimension attributes $A$ is all query columns $q \in Q$ since a column from result tables might not have aligned columns in the query table;
Meanwhile, the measurement attributes $M$ are all the categorical and numerical columns in query and result tables.
A valid visualization plan should satisfy two requirements: (i) There should be no overlap between dimension and measurement attributes, i.e., $A \neq M$ and $M \notin C(A)$; (ii) The result of grouping values in $A$ by $M$ should not be empty.
For the aggregates $F$, we consider the common ones such as COUNT, MIN, MAX, AVG, and SUM, among others.
Following this route, a straightforward solution is to enumerate all plans and compute each plan's utility score.
Then, the result will be plans with top-$n$ highest utility scores.

\subsubsection{Avoid Redundant Computation}

However, such a linear scan is rather expensive due to each plan's overhead of group-by operations.
Since a group-by operation is needed to walk through the tables associated with columns in $A$ and $M$, its cost is related to the table size, which might be very large in the data lake scenario. 
To address this issue, we need to avoid redundant computation in computing group-by.
We observe that when grouping by a given dimension attribute $A$, the result of COUNT aggregation is the same for all measurement attributes $M$.
Thus, when enumerating the plans, if the COUNT aggregation is already calculated for a previous plan group by $A$, we can share the results for all $M$.
Besides, the results of AVG aggregation could be obtained from those of SUM and COUNT ones for the same $A$ and $M$ and do not need a redundant computation.

\subsubsection{Bound Estimation with Sampling}

We can optimize further by reducing the volume of data accessed during the candidate generation process.
To reach this goal, we borrow the idea from previous works in Approximate Query Processing (AQP)~\cite{DBLP:conf/eurosys/AgarwalMPMMS13}.
The high-level idea of AQP is to first execute queries on a small sampled subset of the whole dataset and quickly get an initial answer.
Then, the size of sampled dataset grows gradually, and the query results become more and more accurate.
The whole process will terminate when the budget for accessed data is reached or a certain accuracy guarantee is satisfied.
In our problem setting, when computing the group-by and utility scores, we can gradually examine a subset of the table instead of using the whole table for computation at one time.
In this process, the temporary utility score computed from a subset of data could also serve as a lower bound of the top-$n$ results: if the utility score of a plan is already lower than such a lower bound, the plan could be pruned without computing it over the full tables.
To reach this goal, we employ the Hoeffding-Serfling inequality~\cite{wasserman2013all} from the domain of statistics to deduce such a bound, which could be formally stated in Theorem~\ref{teo-hs}:
\begin{theorem}\label{teo-hs}
	Let $\bigY = y_1, y_2, ..., y_N$ be a set of values in range $[0, 1]$ with a mean value $\mu$; Let $Y_1,Y_2, ...,Y_m$ be a sequence of random variables drawn from $\bigY$ without replacement. For every $k \in [1, N)$ and a probability $\delta > 0$:
	\begin{equation}
		 Pr [\max\limits_{k \leq m < N}  \lvert \frac{1}{m} \sum_{i=1}^m Y_i - \mu \rvert \geq \epsilon_m] \leq  \delta
	\end{equation}
\end{theorem}
where $\epsilon = \sqrt{\frac{(1- \frac{m-1}{N})(2\log \log m+ \log \frac{{\pi}^2}{3\delta})}{2m}}$
 
The core idea of this theorem is the bias between results computed over data samples and the true result is within a given interval with size $\epsilon$, which is related to the number of samples ($m$ in the theorem).
The more samples there are, the smaller the value $\epsilon$ will be, which means the results will be more accurate.
Following the practice of hypothesis testing, the value of $\delta$ is set as 0.05.
In our problem setting, each $Y_i$ above could be regarded as estimating the utility score of a given visualization plan $P$.
Then the value of $\epsilon$ can be computed based on Theorem~\ref{teo-hs}.
Suppose the estimated score is $D$, then the lower (upper) bound would be $D-\epsilon$ ($D+\epsilon$).

\SetKwInOut{Parameter}{Variables}
\begin{algorithm}[!t]
	\small
	\KwIn{The query column $q$; Result tables $\bigS$; The number of results $n'$}
	\KwOut{The set of candidate plans $\bigH$}
	Split $q$ and tables $\bigS$ into several batches $\bigB$\;
	Identify the potential set of visualization plans $\bigP$\;
	Initialize $\bigH = \emptyset$\;
	\ForEach{$B \in \bigB$}{
            $\bigH = \emptyset$\;
            $L_{overall} \leftarrow$ the min lower bound of top $n'$ plans in $\bigH$, initialize to inf\;
		\ForEach{$P \in \bigP$}{
			Compute $P.F$ over $P.A$ and $P.M$ over $B$, resue the COUNT results when necessary\;
			$P.score \leftarrow \bigD(P_B)$;\ // $\bigD(P_B)$ is the partial results computed on $B$\\
			Compute the interval size $\epsilon$\;
                $P.L = \bigD(P_B) - \epsilon$\;
                $P.U = \bigD(P_B) + \epsilon$\;
                \uIf{$|\bigH| < n' $}{
			    Add $P$ into $\bigH$\;
                    $L_{overall} = min(L_{overall}, P.L)$\;
			}
			\Else{
				\uIf{$P.U< L_{overall}$}{
					  $\bigP.remove(P)$\;
				}
                    \Else{
                        Add $P$ into $\bigH$\;
                        \uIf{$P$ is in the top $n'$}{
                            $L_{overall} = min(L_{overall}, P.L)$\;
                        }
                    }
			}
		}
            \uIf{$|\bigH| = n'$}{
			\Return $\bigH$\;
		}
	}
	\Return $\bigH$\;
	\caption{Candidate Generation}
	\label{alg:cands}
\end{algorithm}

\subsubsection{The Pruning Algorithm}

Although the high level idea of Theorem~\ref{teo-hs} has been employed in previous studies about scalable data visualization~\cite{DBLP:journals/pvldb/KimBPIMR15,DBLP:journals/pvldb/MackeZHP18,DBLP:journals/pvldb/RahmanAKBKPR17,DBLP:journals/pvldb/VartakRMPP15}, their solutions cannot be directly applied into our problem.
To realize above idea in our problem setting, we propose an effective pruning strategy as shown in Algorithm~\ref{alg:cands}.
For each column in the query table, it first splits itself and its all associated tables into batches with random shuffles (line: 1) and apply one batch in the computation at one time (line: 4).
Since the results between batches might change drastically, it restarts the ranking in the beginning of each batch by clearing the heap (line: 5).
To facilitate the pruning, a global lower bound is maintained for the top-$n'$ plans in the heap (line: 6). 
For each plan in $\bigP$, the new utility score is calculated by taking the average of all past scores plus the current one for this plan. 
It then computes the interval $\epsilon$ and deduces the lower and upper bound in the plan. 
If the heap size has not reached $n'$ yet, it simply adds the plan and updates the global lower bound (lines 13-15). 
Otherwise, depending on whether the upper bound of $P$ is smaller than the lower bound of top $n'$ in the heap, it decides to discard $P$ (line: 17-18) or add it to the heap (line: 20). 
If $P$ has a score that gets it into the top $n'$ in the heap, then we need to consider updating the global lower bound (lines 16-22). 
Finally, when finishing looping through all plans in $\bigP$, if there are exactly $n'$ plans left in the heap, the process is terminated (lines 23-24).

The pruning methods introduced in Algorithm~\ref{alg:cands} might involve false negative because (i) the estimation is with a probability $\delta$; (ii) the batch shuffling process involves randomness that might discard promising plans with low utility score in earlier batches.
It could be considered as a trading result quality for execution time.
We will later show in Section~\ref{subsec-res} that the proposed solution could make a reasonable trade-off between them.
Here the hyper-parameter $n'$ decides the number of candidates to be verified.
We set it as $n$ empirically in our implementation. 
We will apply the above approach on all the $u$ query columns in the query table and obtain $q*n'$ candidates in the candidate generation step.
Finally, in the Result Ranking step, we will verify each candidate's true value and select the top-$n$ highest results.

\begin{figure}[h]
	\centering
	\hspace{-1.15em}\includegraphics[width=0.4\textwidth]{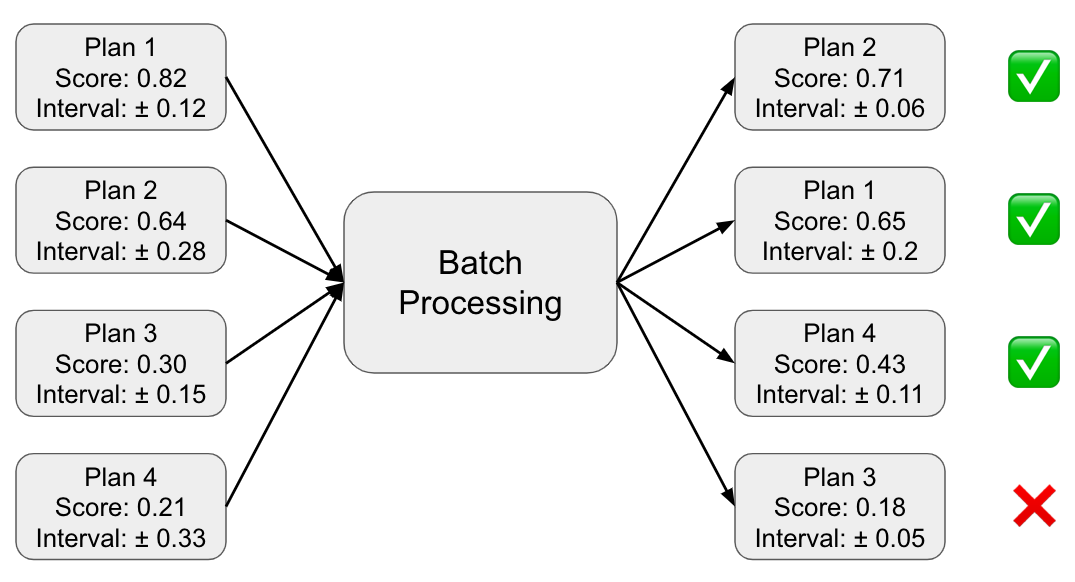}
	\caption{A Running Example of Pruning Process}
	\label{fig:prune}
\end{figure}

\begin{example}
We provide a running example in Figure~\ref{fig:prune} to illustrate how Algorithm~\ref{alg:cands} works.
Suppose there are 4 plans in the second batch ranked in the descending order of estimated scores, and their scores and intervals are shown on the left side.  
After batch processing, their scores and intervals are updated, and so are their rankings. 
If we would like to look for top 2 plans, then plan 2 and 1 are under consideration. 
Plan 4 is still kept because its upper bound (0.54) overlaps with the lower bound of plan 1 (0.45). 
However, Plan 3 can be pruned because its upper bound of 0.23 does not overlap with the lower bound of 0.45 of Plan 1. 
\end{example}

\section{Discussion}\label{sec-disc}

In this section, we discuss the potential opportunities of expanding the functionalities of our proposed framework.

To begin with, now \name presents all results in the format of bar charts.
Nevertheless, the proposed visualization recommendation techniques are not limited to a certain kind of chart.
As visualizations in \name can be represented as a collection of two column tables of the form $\langle a, F(M) \rangle$, similar to~\cite{DBLP:journals/pvldb/VartakRMPP15}, our proposed approach can be easily extended to other visualization types such as line charts and heatmaps.
Regarding the choice of chart type, e.g. decide whether to use line or bar chart for a visualization plan, we can simply follow the routine of previous studies~\cite{DBLP:conf/sigmod/TangWL19}.

In addition, we are aware that there are potential better design choices for some building blocks of our system.
Take the step of series creation as an example, we can obtain a perfect results if there are additional external knowledge about the tables.
For instance, if we know from the metadata that each result table is about the records in a year of a city, we can put tables about the same city into one series.
Then the visualization will present the results of each city from multiple years aggregated from several result tables.
However, consider the inherited characteristics of data lake, e.g. lack of high-quality metadata and inconsistent schema, such external knowledge might not always be available.
Our proposed data-driven approach could serve as a reasonable solution in general scenarios.
Nevertheless, the proposed \name framework is flexible and it is easy to integrate user defined approaches to specify the dimension attributes.
Moreover, regarding the process of handling textual columns, there are also many alternative solutions to define the visualization and obtain insightful plans.
Since our framework do not create extra obstacles of integrating text visualization techniques~\cite{DBLP:journals/tvcg/LiuWCDOEJK19,DBLP:journals/cgf/KucherPK18}, it would be a promising direction of future work to improve the visualization of textual columns with more advanced visualization techniques in our framework.


\section{Evaluation}\label{sec-exp}

\subsection{Experiment Setup}\label{subsec-setup}

\begin{table}[ht]\scriptsize
	\centering
	\caption{The statistics of datasets}\label{tbl-stats}
	\vspace{-2em}
	\begin{tabular}{l|cccc}
		\toprule
		Dataset & \# Query Table & \# Data Lake Table & Avg \# Row per Table  & Size (GB) \\
		\midrule
		TUS & 100 & 5,044 &  2,346 & 1.5 \\
		SANTOS & 80 & 11,086  & 7,706 & 11 \\
		LakeBench & 3,171 & 4,028  & 119,281 & 21 \\
		\bottomrule
	\end{tabular}
\end{table}

\subsubsection{Datasets}

We conduct experiments on three public datasets that have been widely evaluated in previous studies about data discovery.
\tusd~\cite{DBLP:journals/pvldb/NargesianZPM18} is the first benchmarking datasets for table union search.
\santos~\cite{DBLP:journals/pacmmod/KhatiwadaFSCGMR23} is created from Open Data and simulates the real data lake scenarios.
\lakeb~\cite{DBLP:journals/pvldb/DengCCYCYSWLCJZJZWYWT24} is so far the largest benchmarking dataset for unionable and joinable table related tasks with rich manually created ground truth.
Here we use its sub-task of \textsf{OpenData Large}.
The detailed information can be found in Table~\ref{tbl-stats}. 
For each dataset, we run each query with the Starmie~\cite{DBLP:journals/pvldb/FanWLZM23} framework and obtained the top ranked result tables.
Such pairs of query and result tables are considered as the input of our visualization recommendation system.
Note that we just require the knowledge about column alignment between the query table and result tables here but not the specific column-level relatedness scores, which are available for (almost all) existing approaches for data discovery.
Thus other data discovery engine could also be applied here.

\subsubsection{Compared methods}

Since there is no previous study for the problem of visualization recommendation for data discovery from data lakes, and as illustrated in Section~\ref{sec-prelim}, it is non-trivial to extend existing approaches of data visualization to our problem, here we just show the performance of each proposed technique in Section~\ref{sec-method}.
Therefore, we include the following methods for comparison: 
\nom is the method that treats each column in the result table as a series, i.e., there is no series creation process.
\overlap is the baseline method of series creation based on syntactic similarity introduced in Section~\ref{subsec-series};
\stats is the data-driven approach for series creation introduced in Section~\ref{subsec-series};
\prune is the method of applying the pruning techniques in Section~\ref{subsec-opt} on the basis of \stats.

\subsubsection{Evaluation metrics}

We will consider both the efficiency and effectiveness of each compared method introduced above.
We use average execution time per query as the metric for efficiency;
For effectiveness, we make the evaluation based on the practice of previous study~\cite{DBLP:journals/pvldb/VartakRMPP15}: we compute the average utility score of the top-$n$ results of all queries in a dataset.
And the higher the average utility score is, the more effective the compared method is.
For the general usability of our solution, we will further illustrate in the user study later in Section~\ref{sec-user}.

\subsubsection{Environment}
We implemented all algorithms with Python.
The experiments were run on a server with 1 AMD EPYC 7R32 48-core processor and 192GB RAM.
We ran all experiments 5 times and reported the average performance.
We will fix the number of returned visualization plan $n$ as 10 by default.

\subsection{Results}\label{subsec-res}

\begin{figure*}[ht]
	\centering
		\includegraphics[width=0.33\textwidth]{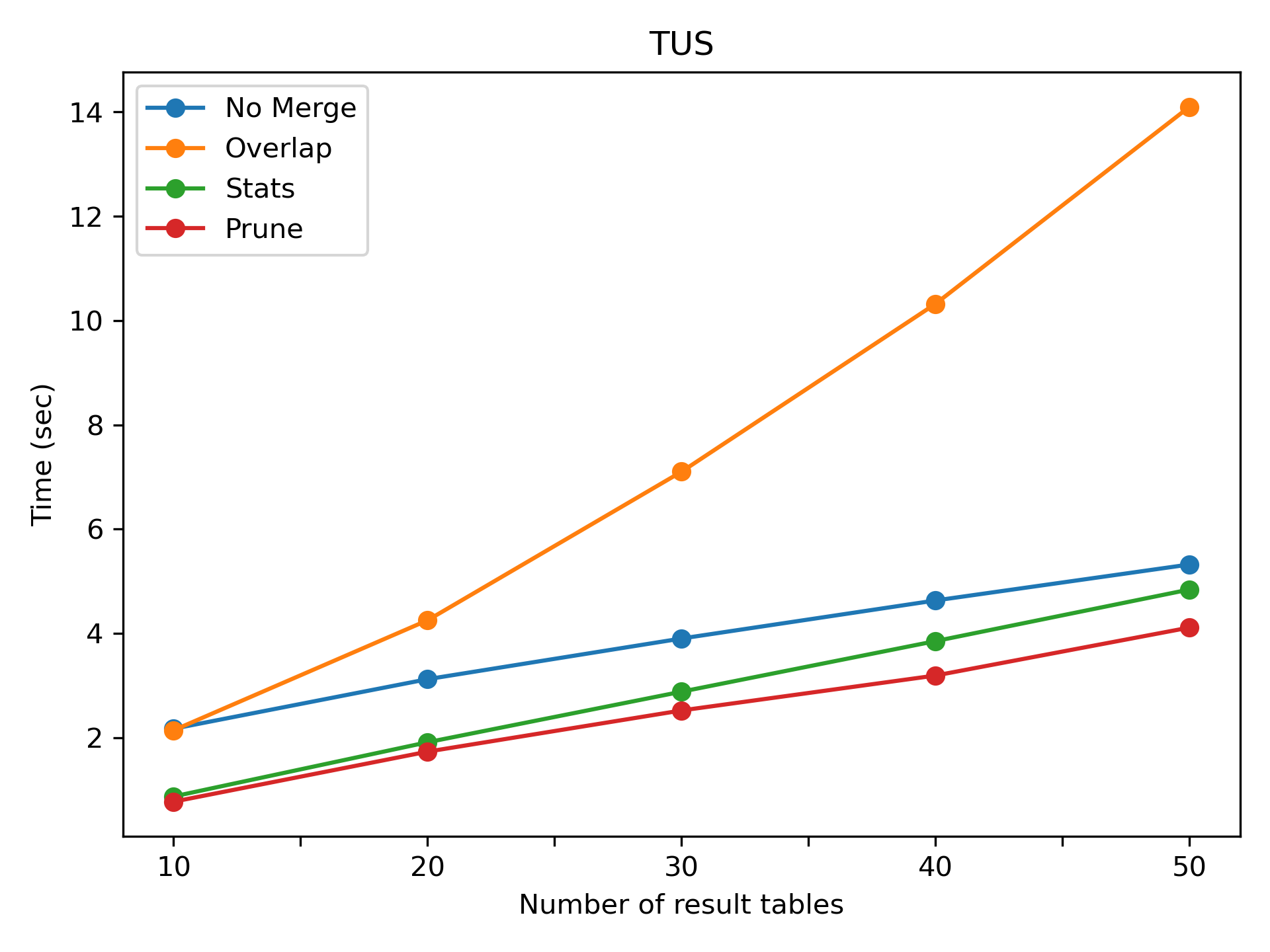}
		\includegraphics[width=0.33\textwidth]{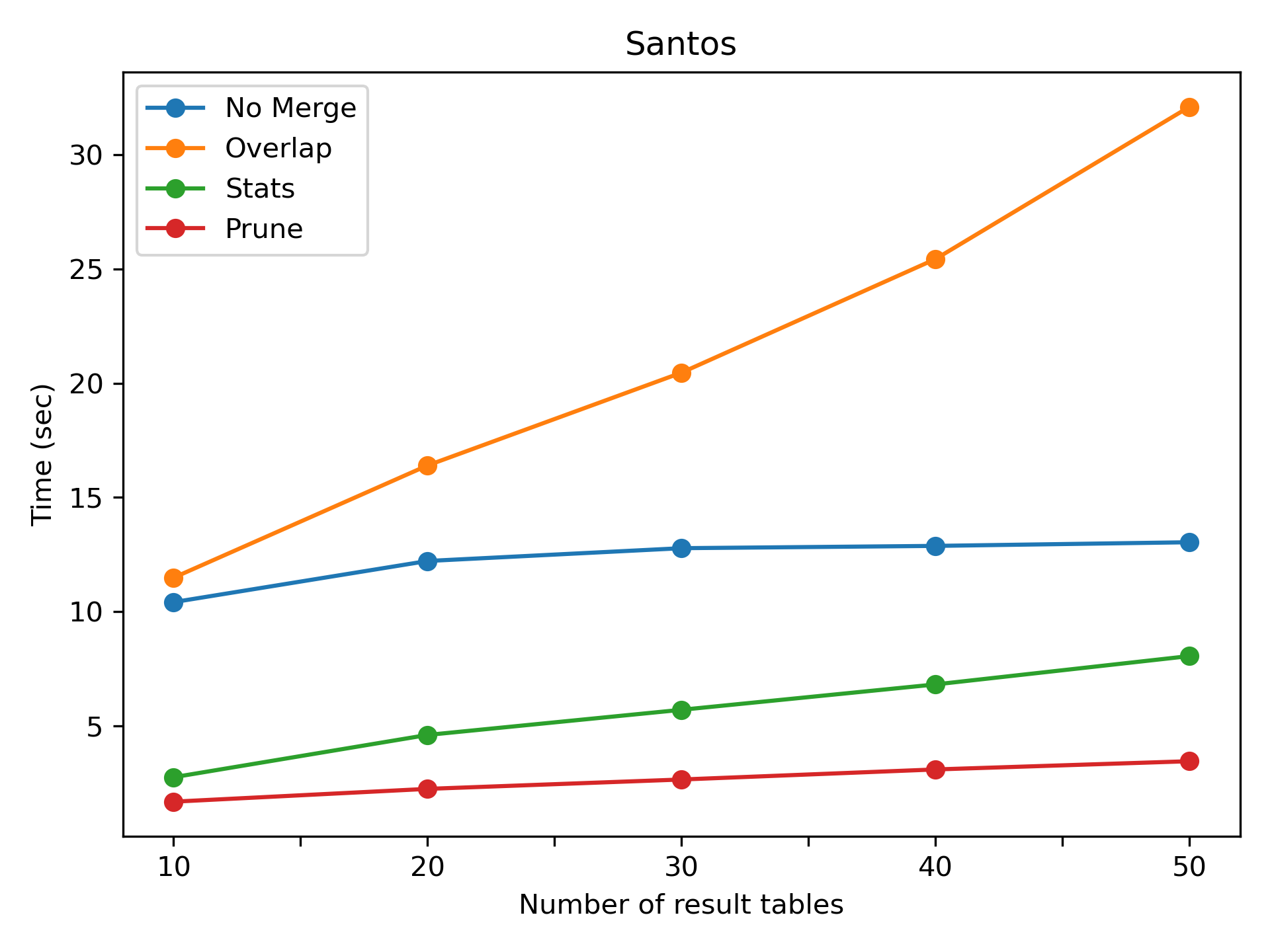}
		\includegraphics[width=0.33\textwidth]{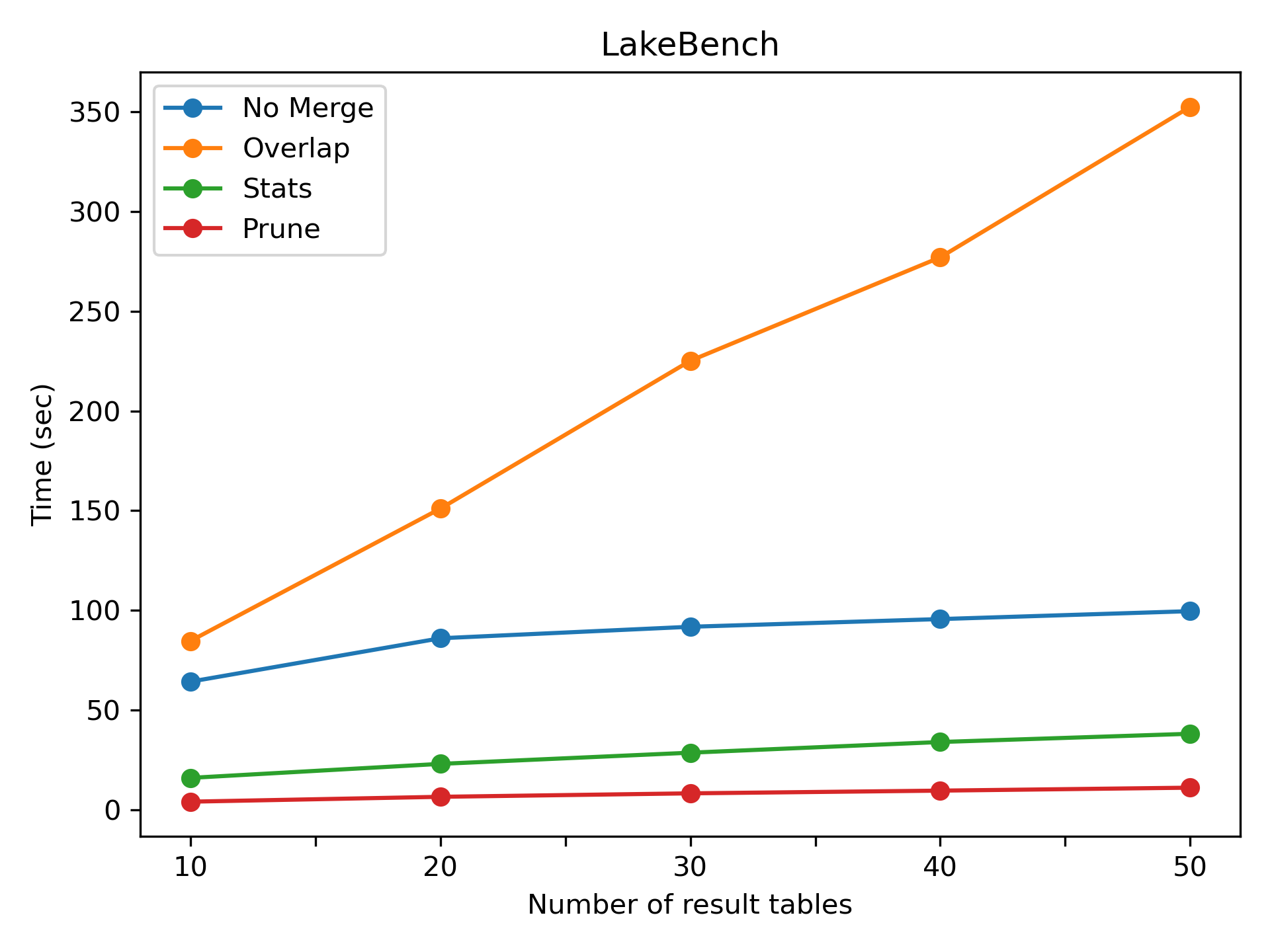}
	\vspace{-2em}
	\caption{Results of Proposed Techniques: Execution Time}
	\label{fig:time}
\end{figure*}

First, we report the results of the execution time of all proposed methods in Figure~\ref{fig:time}.
In this experiment, we vary the number of result tables $k$ in data discovery from 10 to 50 and observe the performance of each method.
And we have the following observations:
Firstly, \stats and \prune achieve better overall performance under all settings, which illustrated the necessity of techniques in the process of series creation and candidate generation.
Specifically, \prune outperforms other compared methods by up to 30 times.
For instance, in the \lakeb dataset when $k = 40$, the average execution time per query of \prune is 9.59 seconds, and while that for \stats, \overlap and \nom is 33.98, 277.09 and 95.64 seconds, respectively.
Secondly, \overlap always performs worst among all methods.
The reason is that it takes a very long time to traverse the columns to compute the syntactic similarity, while \stats only need to estimate and check some statistics.
Besides, the execution time of \overlap increases sharply along with the number of result tables, which is also due to the need to traverse columns.
Thirdly, the gap of performance between different methods on \tusd is relatively small, while that on the \lakeb is more obvious.
This is because as shown in the analysis in Section~\ref{subsec-opt}, the performance of algorithms is closely related to the average number of rows in the query and result tables. 
Since the tables in \tusd are generally not so large as illustrated in Table~\ref{tbl-stats}, the bottleneck in efficiency is not as obvious as the other two datasets.
Lastly, the execution time is relatively constant for all methods but \overlap.
This is because \nom does not involve computation of series creation, while the complexity of \stats and \prune is independent of the total number of rows of all result tables.

\begin{figure*}[ht]
	\centering
	\includegraphics[width=0.33\textwidth]{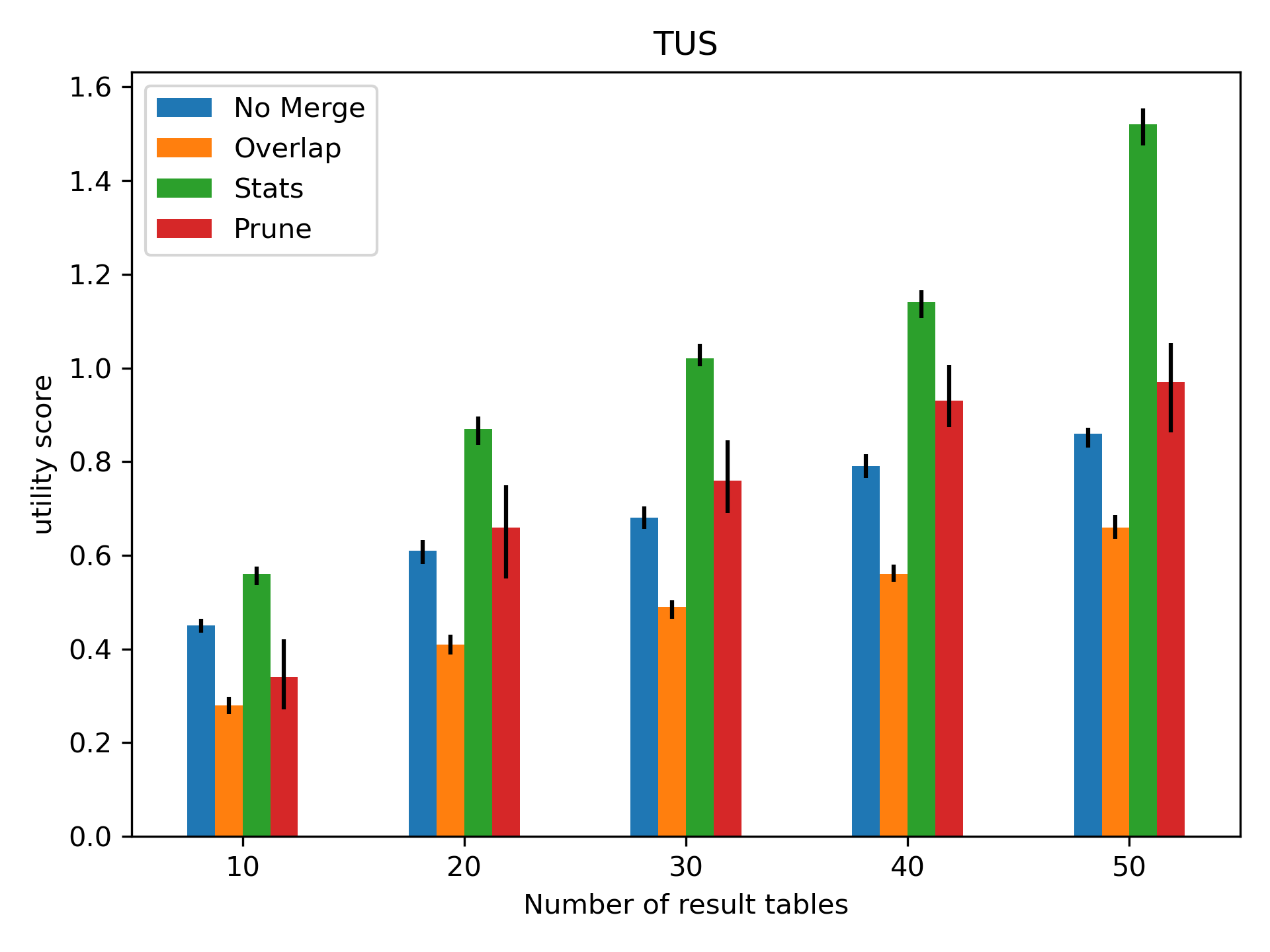}
	\includegraphics[width=0.33\textwidth]{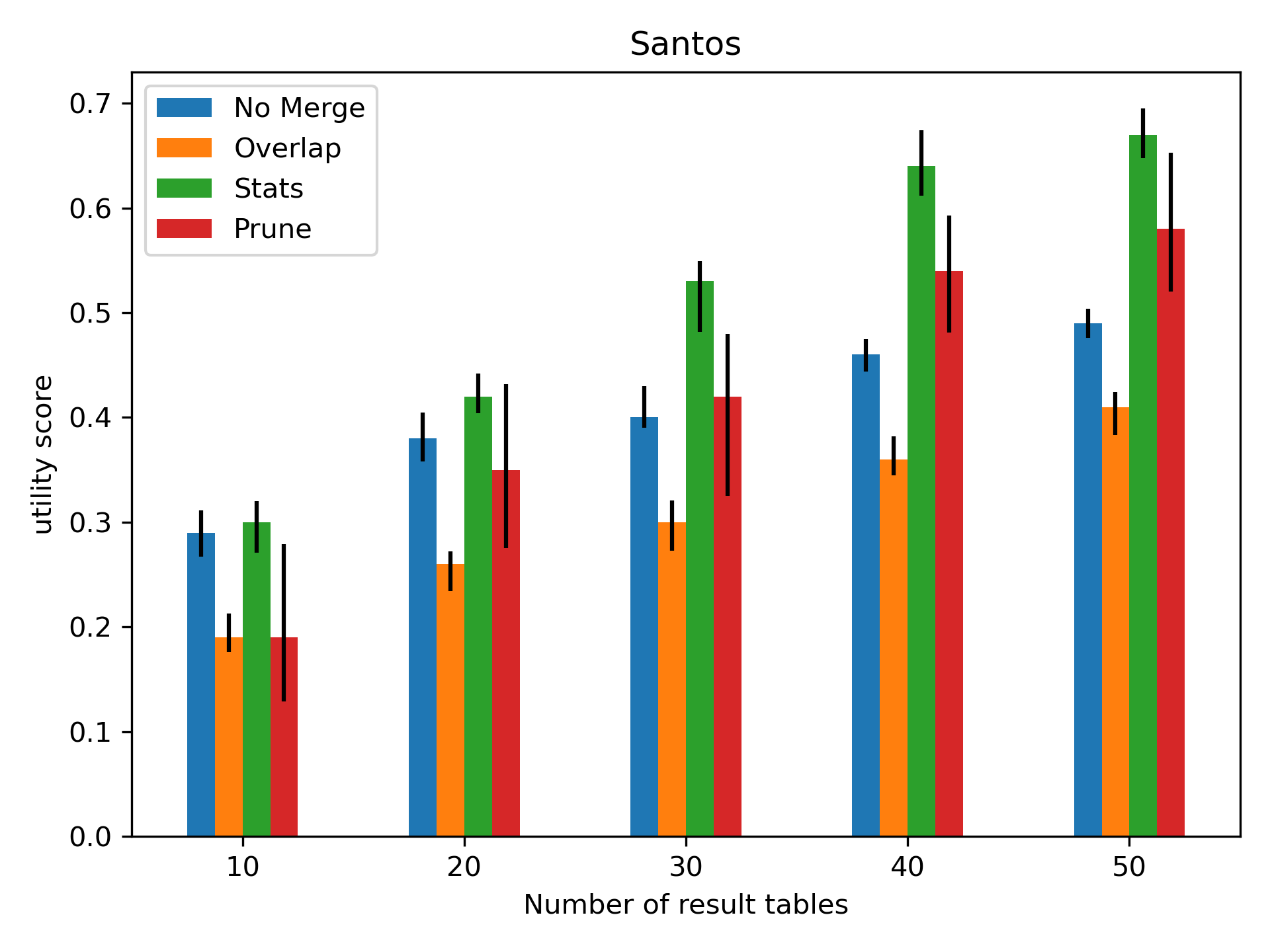}
	\includegraphics[width=0.33\textwidth]{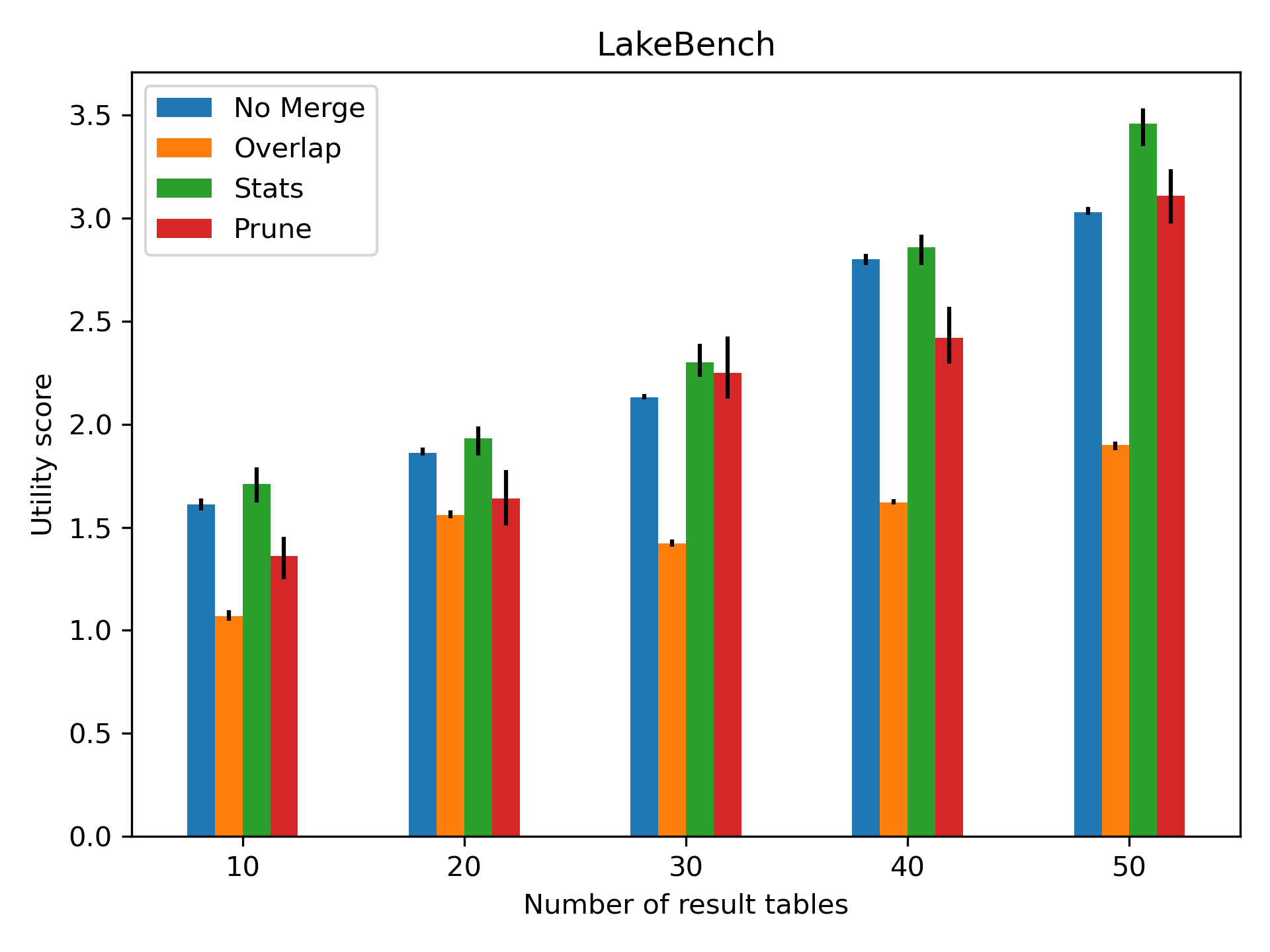}
	\vspace{-2em}
	\caption{Results of Proposed Techniques: Effectiveness. The error bar shows the lower bound and upper bound of results for methods with randomness.}
	\label{fig:effectiveness}
\end{figure*}

Then we show the effectiveness of all proposed methods.
As discussed in Section~\ref{sec-method}, it is not realistic to obtain the optimal overall utility score due to the exponential search space.
Thus instead of computing the recall or accuracy, we evaluate the quality of top-$n$ results by looking at the average utility score of them.
Since both \stats and \prune relied on statistical methods that involve randomness, we also include the error bars of these methods to denote the lower and upper bounds of them in multiple runs.
The results are shown in Figure~\ref{fig:effectiveness}.
We can see that generally speaking \stats has the overall best results in effectiveness, which achieves the largest scores under all settings.
For example, on the \santos dataset when $k = 30$, the average utility score of \nom, \overlap, \stats, and \prune is 0.4, 0.3, 0.53 and 0.42 respectively.
Note that since \nom always has a larger number of bars than \stats and \prune, the total utility score tends to be larger based on the way to calculate it as shown in Equation~\ref{eq-emd}. 
Actually the actual performance of \stats and \prune will be even more outstanding, considering the penalty regarding perceptual
scalability brought by the number of bars in \nom. 
Here, we did not apply such penalties in the utility score to make it consistent with the previous study~\cite{DBLP:journals/pvldb/VartakRMPP15}.
The pruning strategies developed in Section~\ref{subsec-opt}  involved some false negatives that might discard some promising visualization plans.
Therefore, we can consider \prune makes a trade-off between effectiveness and efficiency; thus its effectiveness results are worse than \stats.
Besides, there is also an observation that the advantage of \stats becomes more obvious when there is a larger number of result tables.
This is because \stats is a data-driven approach, and more data could result in more accurate estimation.
Moreover, we also see that \nom performs generally better than \overlap since syntactically similar columns do not necessarily lead to related columns.
It also shows that simple heuristics cannot easily handle the series creation step, and it is essential to introduce the data-driven approach, as we did.

\begin{figure}[ht]
	\centering
	\includegraphics[width=0.48\columnwidth]{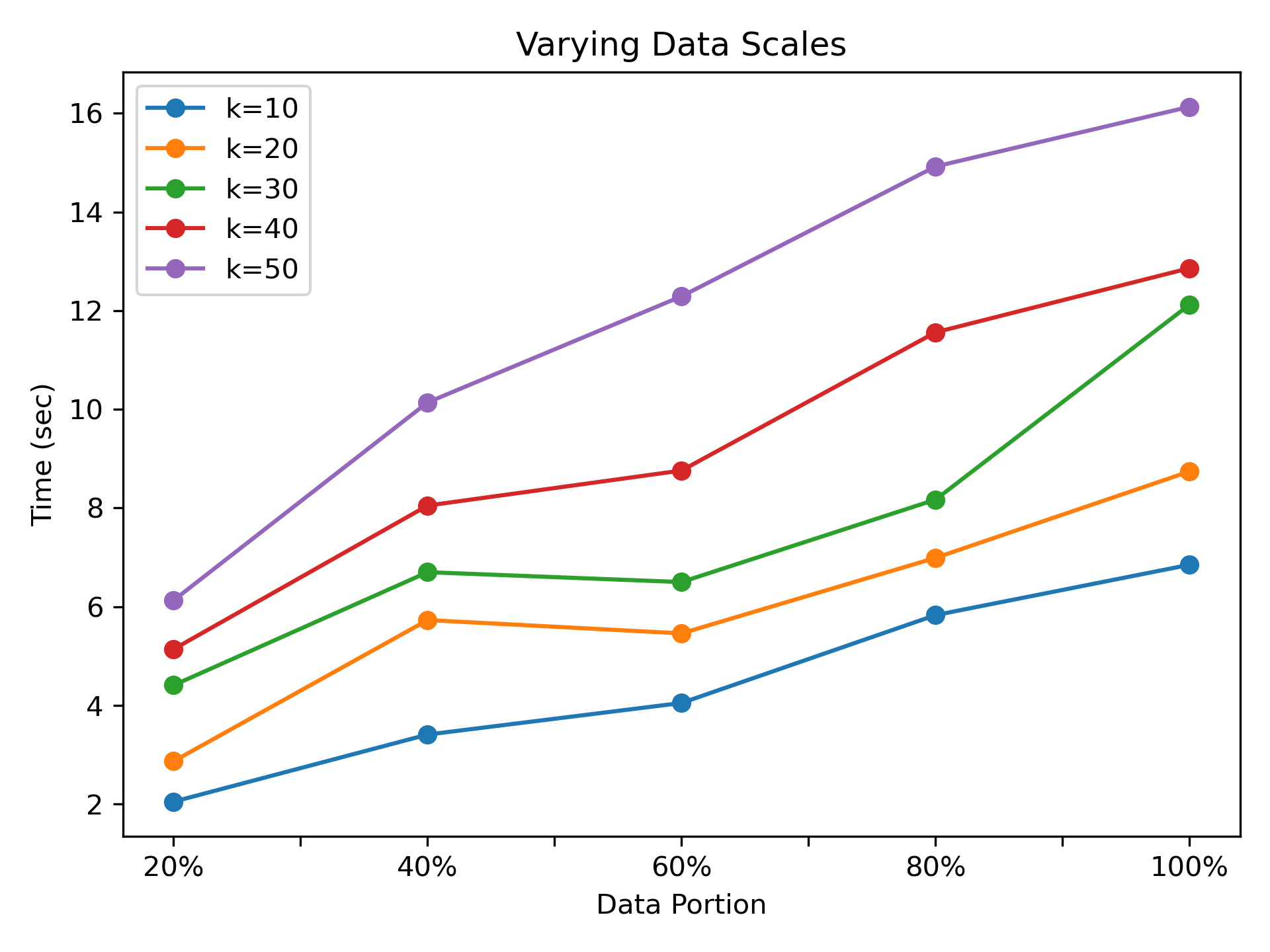}
	\includegraphics[width=0.48\columnwidth]{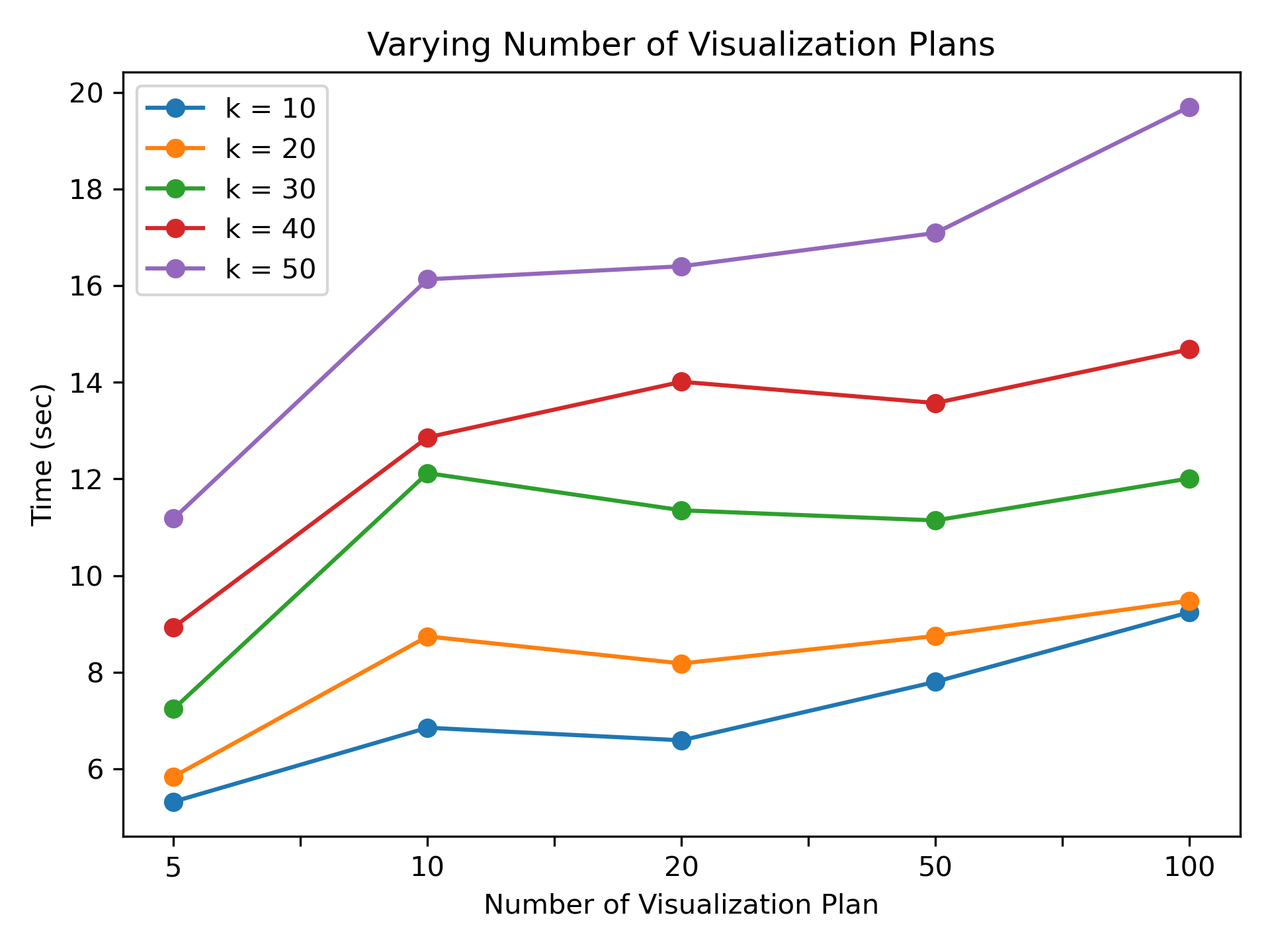}
	\vspace{-1em}
	\caption{Results of Scalability on LakeBench dataset}
	\label{fig:scale}
\end{figure}

Finally, we also investigate the scalability of our \name framework  (i.e. above \prune method) w.r.t. other two important factors:  the data size of involved tables and the number of returned visualization plans.
Since \lakeb is much larger than the other two datasets, we only conduct scalability experiments on it.
To evaluate the scalability w.r.t. data scale, we select the top-20 queries with the largest number of rows in query and result tables from \lakeb.
For these queries, we vary the data scale from 20\% to 100\% for all the query and result tables and observe the performance of our proposed method.
As shown in the left sub-figure of Figure~\ref{fig:scale}, \name achieves near linear scalability under all settings.
Then we vary the number of returned visualization plans $n$ and show results in the right sub-figure of Figure~\ref{fig:scale}.
We can see that the execution time stays stable before $n$ reaches 50 and sometimes larger $n$ even leads to less time.
The reason is that when doing the group-by operations to generate candidates, many plans can have the same dimension attributes but different measurement attributes; thus, they have the same utility score.
When the number of $n$ increases, the utility score of the top element on the min heap might not change, and our search algorithm converges at a similar time or even faster due to the shared computation of aggregation developed in Section~\ref{subsec-opt}.
When $n$ is larger than 50, we observe that the utility score of the top element also increases greatly, and thus, it takes more time to finish searching.

\subsection{Case Study}\label{subsec-case}

\begin{figure}[h]
	\centering
	\hspace{-1.15em}\includegraphics[width=0.45\textwidth]{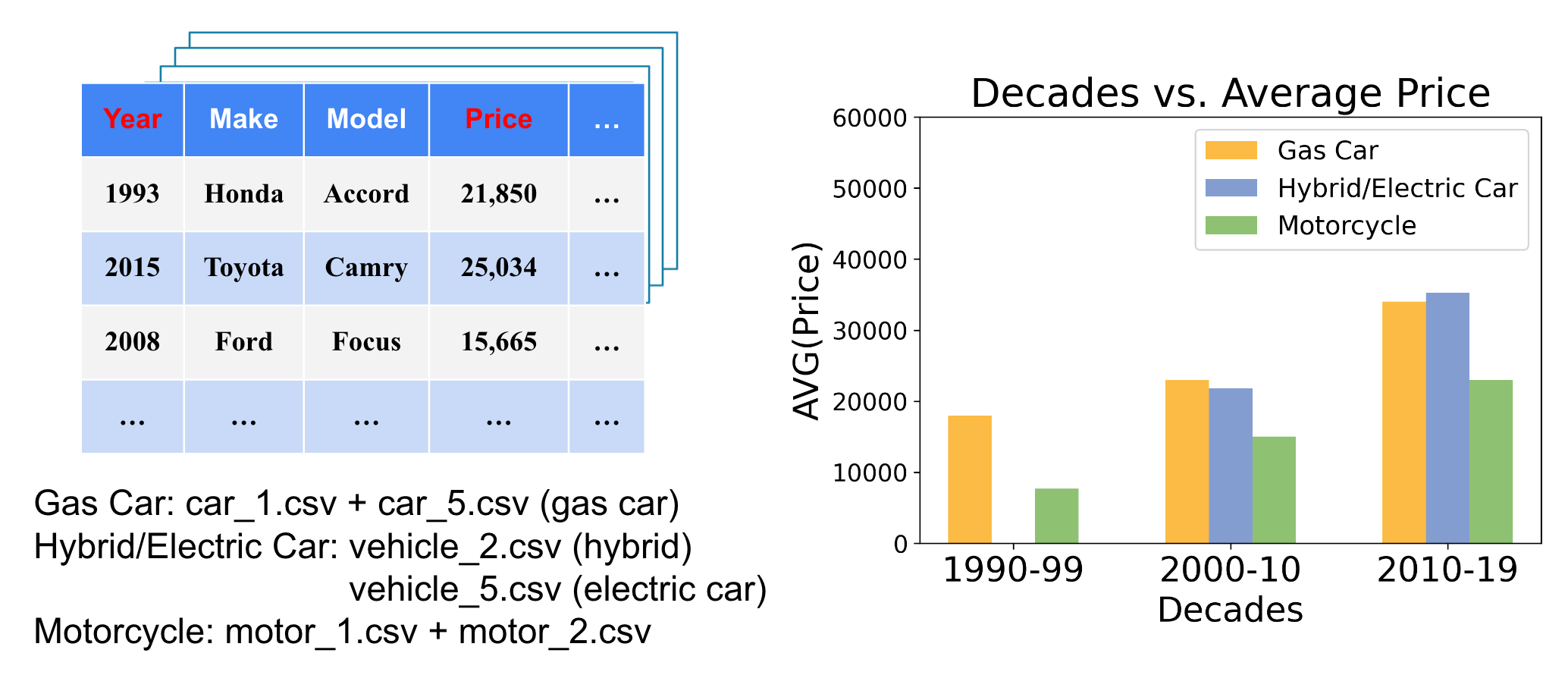}\vspace{-.5em}
	\caption{Case Study:the series created based on distribution of Year shows an interesting trend about fuel of cars.}
	\label{fig:case1}
\end{figure}

Next we show two case studies about some interesting visualization plans recommended by our proposed system.
Our goal is to illustrate the useful insights of data discovery in the visualizations brought by proposed techniques.
First, we would like to illustrate an example of series creation in addition to the data discovery results.
Figure~\ref{fig:case1} shows an example with a query table and result tables about cars.
This example is about a car sales scenario that presents information such as the price, make, and date of the cars.
In this visualization chart, the dimension attribute $A$ is a numerical column \emph{Year} which is divided into several bins as introduced in Section~\ref{subsec-sys}; the measure attribute $M$ is a numerical column \emph{Price} and the aggregate $F$ is AVG.
Specifically, the series are created based on the data distribution of the \emph{Year} column in query and result tables.
We find that the created series could illustrate a correlation with the fuel of cars~\footnote{The labels of series are manually created by us based on the available metadata of tables.}.
As a result, visualization with the above series could also deliver the information that the number of Hybrid/Electric cars has grown much faster than Gas cars in the last 10 years.
It would save the efforts of data scientists to discover it by manual programming. 

\begin{figure}[ht]
	\centering
	\hspace{-1.15em}\includegraphics[width=0.45\textwidth]{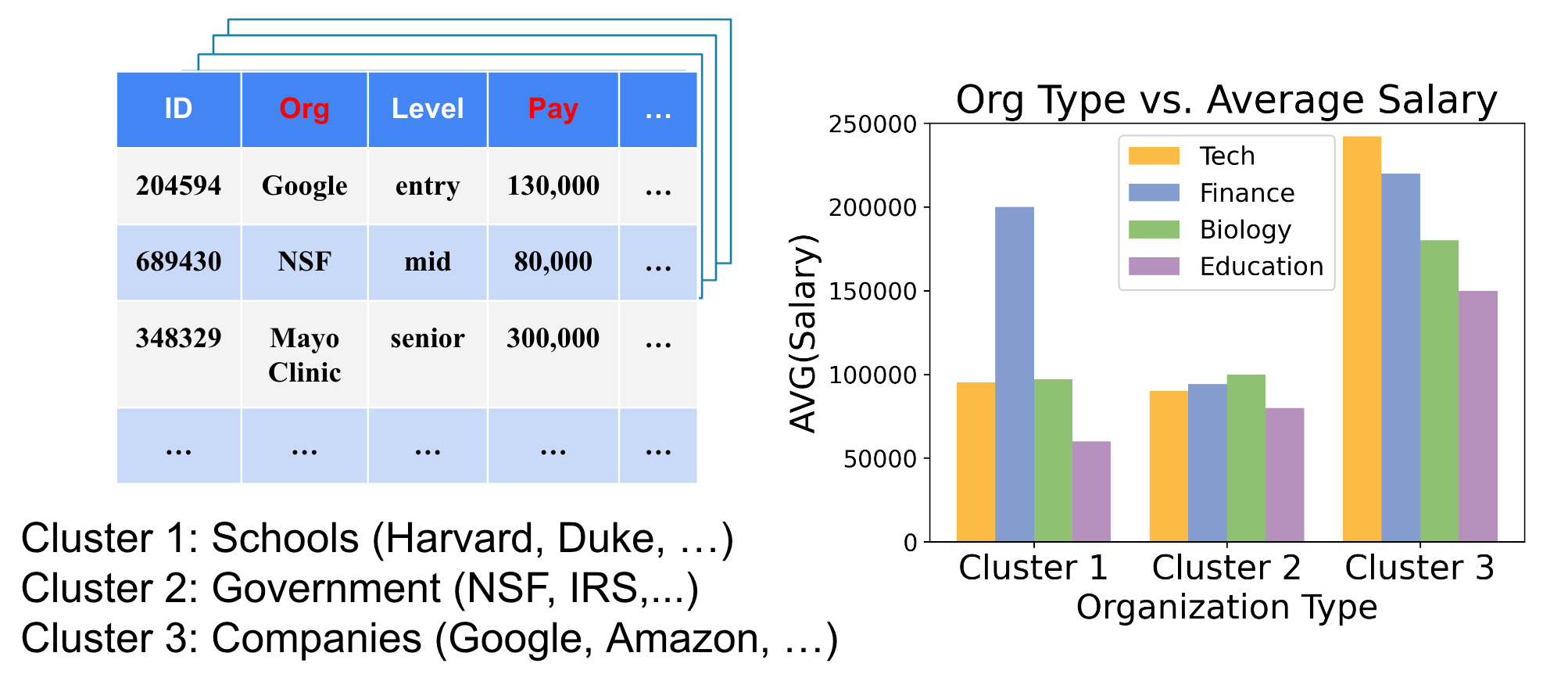}\vspace{-.5em}
	\caption{Case Study: A Visualization Plan with Textual Columns as Dimension Attribute.}
	\label{fig:case2}
\end{figure}

Another important case we want to show is about the situation where textual columns serve as the dimensional attribute $A$ in a visualization plan.
As discussed before, an important difference between \name and previous data visualization works for relational databases is that our work needs to handle the textual columns and recognize visualization plans composed with them.
Figure~\ref{fig:case2} shows an example with a query table about employee information, and result tables with similar contents.
And \name created a visualization plan where $A$ is a textual column \textsf{Organization}, $M$ is \textsf{Salary} and $F$ is AVG.
Here, the Organization columns consist of names of different organizations where employees work.
Our approach splits the cells of such columns into several clusters, and each value of the dimensional attribute corresponds to a cluster, as shown in the bottom left of Figure~\ref{fig:case2}.
Here, we find the latent semantics of the clustering results: each cluster corresponds to a kind of organization, i.e., school, government, and company.
Then this visualization plan could further illustrate the average salary in different kinds of organizations rather than a specific one.
We can see that even the simple heuristics in Section~\ref{subsec-sys} could help discover such interesting visualizations.

\section{User Study}\label{sec-user}


\begin{table*}[t]
\scriptsize
\centering
\caption{\small Example quiz tasks and their corresponding purposes and actions.}
 \vspace{-10pt}
\begin{tabular}{p{8cm} p{9cm}}
\hline
Questions & Purpose and actions   \\ \hline
\textbf{Q1}: List all primary colors covered in the result tables. & \code{browse} tables $\rightarrow$\code{lookup} column $\rightarrow$\code{generate} set\\
\textbf{Q2}: Which animal type appears as the highest frequency in the most number of states?  &  \code{browse} tables $\rightarrow$\code{lookup} column $\rightarrow$\code{generate} aggregate$\rightarrow$\code{compare}\\  
\textbf{Q3}: What is the third most frequent status tag across all tables?  & \code{browse} tables $\rightarrow$\code{lookup} column $\rightarrow$\code{generate} aggregate$\rightarrow$\code{compare}\\ 
\textbf{Q4}: What is the average price of rabbits across all tables?  &  \code{browse} tables $\rightarrow$\code{lookup} column$\rightarrow$\code{identify} subset  $\rightarrow$\code{generate} aggregate\\ 
\textbf{Q5}: What is the min gross cost of the records with 10mg strength from the state of Florida? & \code{browse} tables $\rightarrow$\code{lookup} column $\rightarrow$\code{identify} subset$\rightarrow$\code{generate} aggregate $\rightarrow$\code{compare}\\ 
\textbf{Q6}: What is the second least frequent presentation among results in the year 2014?&  \code{browse} tables $\rightarrow$\code{lookup} column $\rightarrow$\code{identify} subset$\rightarrow$\code{generate} aggregate $\rightarrow$\code{compare}\\ 
\end{tabular}
\label{tab:questions}
 \vspace{-5pt}
\end{table*}

\subsection{Design}\label{subsec-design}

We compare \name with literate programming tools such as computational notebooks typically used by data scientists for exploring data discovery results in data lakes. 
Our study was designed to answer the following questions: 
\textbf{RQ1)} \emph{How does an interactive system like \name impact the usability and efficiency of exploring data discovery results?} 
and \textbf{RQ2)} \emph{What are the strengths and limitations of \name in terms of user experience compared to literate programming tools? }

\stitle{Datasets and quiz tasks.} We selected two query tables from the popular Open Data repository~\footnote{https://www.data.gov} and retrieved the top-$10$ unionable tables returned by Starmie~\cite{DBLP:journals/pvldb/FanWLZM23} as data discovery results.  
The two query tables were related to pet and medical domains which could represent real world tasks for data scientists.
The number of columns in the two query tables is 11 and 12, respectively.
Table~\ref{tab:questions} lists the questions provided to the participants when using computational notebooks. 
Meanwhile, to make a fair comparison, questions for the visualization systems are similar in difficulty but with different attributes and predicates. 
These questions were designed to introduce an increasing level of difficulty to the participants. 
Therefore, we did not randomize the order of the questions. 
These questions represent a range of data analysis tasks that users typically perform over data discovery results (see supplementary material for details.) 
We leverage the typology of data exploration tasks~\cite{DBLP:journals/tvcg/BrehmerM13} to identify the purposes and actions required for each question. 
This typology encompasses a range of domain-independent tasks on visual data representations and has been widely utilized as a guideline for developing models of visualization systems and defining the scope of tasks in various domains, e.g., interactive task authoring, document mining, and multivariate network analysis~\cite{DBLP:journals/pvldb/RahmanBLZSKP21}.
All of these tasks required participants to complete several actions representing one of the following purposes: \code{browse} (searching based on characteristics where location is unknown), \code{lookup} (searching based on entities where location is known), \code{identify} (returning the characteristics of entity found during search), \code{generate/record} (generation or recording of new information), and \code{compare} (returning characteristics of multiple entities).

\stitle{Study participants and phases.} We recruited $12$ participants ($3$ female and $9$ male) for the study via Slack outreach at \company. 
The participants had different technical roles and expertise, such as Research Engineers (principal, senior, and junior), Research Scientists (senior and junior), and Full-stack developers. 
The study was conducted by a co-author of the paper. 
And none of the paper authors were study participants. 
The study consisted of three phases: introduction, quiz, and survey.
In the introductory phase, participants were provided a brief overview of the study objectives and follow-up phases. 
The quiz phase required participants to solve several tasks related to the exploration of data discovery results within a data lake using both a computational notebook by writing Python programs and \name. 
However, we alternated the order of systems between consecutive participants. 
Before answering the quiz with a system, participants were provided with a tutorial to help familiarize themselves with the functionalities  --- for \name, participants were introduced to various features; and for computational notebooks, participants were provided with a suite of utility functions to help in data discovery. 
Following the tutorial, participants performed several warm-up tasks before proceeding to the actual quiz tasks. 
Upon the completion of the quiz phase, participants were provided with a survey via Google Form to gauge their impressions about both systems.

\stitle{Evaluation}. We evaluated the completion time and accuracy for
all of the tasks. 
Moreover, we analyzed the survey responses to quantify the usability of both systems. We also collected qualitative feedback during the survey to understand the benefits and limitations of both systems. In presenting participant responses in
Section~\ref{subsec:survey}, we often included excerpts from complete responses (indicated by $\ldots$). 
Additionally, we corrected spelling and grammatical mistakes. 
Therefore, the quotes presented in this paper are essentially paraphrases.

\subsection{Results and Analysis}\label{subsec-userres}

We analyze the quantitative and qualitative data collected during the quiz and interview phases to address our research questions.

\begin{table}\scriptsize
	\centering
	\caption{Results of User Study. \# Answers means how many users provided answers to the question; Accuracy means the portion of correct answers.}\label{tbl-ures}
 \begin{tabular}{l|c|cccccc}
 \toprule
 \textbf{Metric} & \textbf{Method} & \textbf{Q1} & \textbf{Q2} & \textbf{Q3} & \textbf{Q4} & \textbf{Q5} & \textbf{Q6} \\
 \midrule
 \multirow{2}{*}{Max Submission Time (s)}& Notebook & 336 & 509 & 364 & 613 & 173 & 216 \\
 & \name & \textbf{85} & \textbf{117} & \textbf{88} & \textbf{266} & \textbf{68} & \textbf{145} \\
\midrule
 \multirow{2}{*}{Min Submission Time (s)}& Notebook & 87 & 70 & 42 & 144 & 64 & \textbf{82} \\
 & \name & \textbf{32} & \textbf{20} & \textbf{14} & \textbf{123} & \textbf{36} & 85 \\
 \midrule
 \multirow{2}{*}{Average Submission Time (s)}& Notebook & 209.1 & 194.7 & 191.5 & 319.5 & 119.1 & 163 \\
 & \name & \textbf{54} & \textbf{62.7} & \textbf{46.5} & \textbf{172.8} & \textbf{51.75} & \textbf{111.4} \\
 \midrule
 \midrule
 \multirow{2}{*}{\# Answered}& Notebook & 12 & 12 & 12 & 12 & 5 & 4 \\
 & \name & 12 & 12 & 12 & 12 & \textbf{12} & \textbf{12} \\
 \midrule
 \multirow{2}{*}{Accuracy (\%)}& Notebook & \textbf{100} & 100 & 91.7 & 25 & 41.7 & 25 \\
 & \name & 91.7 & 100 & \textbf{100} & \textbf{100} & \textbf{100} & \textbf{100} \\
 \end{tabular}
 \label{tab:results_accuracy_time}
 \end{table}

\noindent\textbf{Task Submission Performance.}\hspace{.5em}
In Table~\ref{tab:results_accuracy_time}, we show the response times of
participants for each task, using computation notebook and \name, respectively. 
For all the tasks, participants’ submission times using \name were obviously lower than notebook on average. 
Moreover, all of the participants completed at least five tasks in less time using \name. 
In fact, there were only two instances (Q4 by $P9$ and Q6 by $P10$) where participants took more time to complete tasks using \name. \smallskip

\noindent\textbf{Task Accuracy.}\hspace{.5em}
As shown in Table~\ref{tab:results_accuracy_time}, for the easier tasks, Q1, Q2, and Q3, participants exhibited similar accuracy. Somewhat surprisingly, while using notebooks, seven and eight participants did not complete their submissions for the more difficult tasks Q5 and Q6, respectively. These tasks required participants to perform a series of operations on the data lake. However, to effectively implement those operations, participants had to formulate their understanding of the tables, relevant columns, and suitable sub-selections and then manually compare different candidates. Therefore, the errors in completing the eventual task may stem from the cognitive load associated with reasoning over the result of these operations. Even though all the participants attempted Q4 when using a notebook, only 25\% of the participants provided correct responses due to task complexity.  

\vspace{-1em}
\subsection{Participant Feedback}
\label{subsec:survey}

Our analysis of the survey results revealed that $83.3\%$ participants preferred \name to computational notebooks. These participants specifically found \name visualizations useful in synthesizing information across multiple tables. On a scale of 1 to 5 (1= not challenging at all, 5=very challenging), the participants rated \name to be easier to use ($\mu = 2, \sigma = 0.27$) compared to notebooks ($\mu = 3, \sigma = 0.81$). We now discuss the strengths and limitations of both systems with respect to the qualitative feedback provided by the participants. 

\noindent\textbf{Getting started with analytics sessions.}\hspace{.5em} Majority of participants ($N=8$) found \name useful in launching the analytics sessions, specifically when working with unfamiliar datasets or a larger data collection --- \response{$\ldots$ \name provides the quick overview of data which is helpful to decide which direction we should go deeper} ($P4$.) In contrast, participants ($N=7$) exhibited negative impressions towards notebooks when launching a session for the first time due the cumbersome interactions via iterative programming --- \response{you have to write some code before getting insights, might also require more understanding of the schema and values in the tables to write valid functions $\ldots$ takes a bit of trial-and-error $\ldots$} ($P12$.) 

\noindent\textbf{Familiarity and learning curve.}\hspace{.5em} In addition to the challenges in getting started, participants noted additional factors as potential limitations of notebooks such as lack of recall (of appropriate Dataframe operations) and a steeper learning curve (with literate programming.) For example, $P1$ commented --- \response{Notebooks require users to be very familiar with Dataframes, while most users might forget if they don't use (Dataframes) frequently.} On the other hand, participants found visualizations presented in \name intuitive. Three participants mentioned that \name would be very helpful for non-technical users due to the low barrier to familiarizing with visualizations --- \response{ the visualization didn’t need any prior familiarity} ($P11$.) 

\noindent\textbf{Convenience and ease of use.}\hspace{.5em} Participants ($N=7$) also found \name to be easier to use compared to notebooks. In facts, participants obtained insights faster with \name as demonstrated in Table~\ref{tab:results_accuracy_time}. For example, $P12$ commented --- \response{$\ldots$ faster to get insights, visualization tend to be
more informative and provide much more insights than the notebooks.} On the other hand, participants deemed notebook operations to be time consuming --- \response{Using Notebook took more time because I needed coding} ($P2$.) Another participant ($P10$) commented --- \response{$\ldots$ hard to manipulate (notebooks) unless you know (Dataframe) operations well.} 

\noindent\textbf{Flexibility of exploration.}\hspace{.5em} Participants ($N=4$) appreciated the higher flexibility afforded by notebooks, specifically when performing more in-depth analysis. For example, newly observations may prompt an analyst to issue new queries over the data lake on-the-fly. However, visualizations provided in \name are pre-defined and therefore, limits users's degrees of freedom. For example, one participant ($P12$) mentioned that notebooks provided \response{a lot more flexibility in terms of analysis that can be done and analysts who are familiar might find it easier to use than learning to interact with a visualization tool.} Another participant commented --- \response{A notebook-based system enables us to conduct detailed and finer-grained operations if needed} ($P4$.) The same participant found \name to be limiting --- \response{\name sometimes lacks the customizability, especially when users want to execute complicated analysis.} Another participant mentioned --- \response{visualizations can sometimes be limited and less flexible when data is pre-aggregated in the visualization in a way that is far from the desired aggregation} ($P12$.) 

\noindent\textbf{Scope and functionalities.}\hspace{.5em} In terms of explorations goals, participants preferred visualizations to locate extremities in data and performing comparisons --- \response{I would prefer Visualization for the purpose like finding the max/min values} ($P2$.) Participants preferred notebooks for dynamically computing aggregations over attributes of interest. One participant ($P3$) found the top-$K$ visualizations to be overwhelming due to the visual discontinuity caused by scrolling up and down the ranked list and requested features such as performing natural language querying over the visualizations. In fact two other participants requested similar features --- \response{For \name, having a simple search might be helpful. For example: show me any graphs that are related to a data column} ($P8$.) Participants ($N=2$) also commented on the uncertainty introduced by approximations of the top-$K$ results and requested for greater transparency when communicating the results.

\section{Related Work}\label{sec-related}

\subsection{Data Discovery from Data Lakes}\label{subsec-datalr}

There is a long stream of research works about data discovery from data lakes in the data management community.
The examples of data discovery tasks include finding related tables~\cite{DBLP:conf/sigmod/SarmaFGHLWXY12}, schema complement~\cite{DBLP:conf/icde/KoutrasSIPBFLBK21}, domain discovery~\cite{DBLP:journals/pvldb/OtaMFS20} and column annotation~\cite{DBLP:journals/pacmmod/MiaoW23}.
Among them the problem of finding related tables attracts more attentions in the recent years.
There are two sub-tasks in this application, namely joinable table discovery and table union search~\cite{DBLP:conf/sigmod/SarmaFGHLWXY12}.
Joinable table discovery aims at finding tables that can be joined with the given table on a column.
\textsf{LSH Ensemble}~\cite{DBLP:journals/pvldb/ZhuNPM16}  and \textsf{JOSIE}~\cite{DBLP:conf/sigmod/ZhuDNM19} employed syntactic similarity functions to decide joinability.
\textsf{PEXESO}~\cite{DBLP:conf/icde/DongT0O21}  computed the fuzzy similarity between columns based on pre-trained word, and \textsf{Deep Join}~\cite{DBLP:journals/pvldb/Dong0NEO23} relied on fine-tuned BERT model to capture the semantic information of columns.
\textsf{MATE}~\cite{DBLP:journals/pvldb/EsmailoghliQA22} focused on the problem of composited join key with multiple columns.
And \textsf{Juneau}~\cite{DBLP:conf/sigmod/ZhangI20} studied the joinable discovery over semi-structured data.
The bottleneck of Table Union Search is to find a proper way to decide the column unionability scores.
Nargesian et al.~\cite{DBLP:journals/pvldb/NargesianZPM18} proposed the first definition and design space for this problem.
The \textsf{$D^3$L} system~\cite{DBLP:conf/icde/BogatuFP020} divided columns into different categories and computed unionability score accordingly.
\textsf{SANTOS}~\cite{DBLP:journals/pacmmod/KhatiwadaFSCGMR23} utilized the knowledge base to decide unionablity and \textsf{Starmie}~\cite{DBLP:journals/pvldb/FanWLZM23} learned a BERT based encoder for table columns.
\textsf{Gen-T}~\cite{DBLP:conf/icde/FanSM24} tried to reclaim the original tables from unioned ones.

There are also some studies targeting at improving the usability of data lake related tasks.
\textsf{Aurum}~\cite{DBLP:conf/icde/FernandezAKYMS18} provided a declarative query language to express data discovery tasks;
 \textsf{RONIN}~\cite{DBLP:journals/pvldb/OuelletteSNBZPM21} constructed a GUI to support navigation of data lakes~\cite{DBLP:conf/sigmod/NargesianPZBM20}, where users can easily browse the tables in the data lake.
 Auctus~\cite{DBLP:journals/pvldb/CasteloRSBCF21} aimed at improving the usability of keyword search over data lakes while Humboldt~\cite{DBLP:journals/corr/abs-2408-05439} automatically provided customized UIs for data discovery based on meta-data.
 None of above efforts can support our task that generates visualizations for results of data discovery tasks. 

\subsection{Visualization Recommendation Systems}\label{subsec-visr}

Visualization recommendation systems can be broadly categorized into rule-based and machine learning (ML)-based approaches. 
Rule-based systems, such as Voyager~\cite{DBLP:journals/tvcg/WongsuphasawatM16}, SeeDB~\cite{DBLP:journals/pvldb/VartakRMPP15}, and AVA~\cite{DBLP:journals/vi/WangLLPWGLZXDLZLC24}, rely on heuristics derived from expert experience or empirical studies. 
Technically, they shared the similar idea with earlier work of exploring results of OLAP queries~\cite{DBLP:journals/pvldb/JoglekarGP15,DBLP:conf/vldb/Sarawagi00,DBLP:conf/edbt/SarawagiAM98}.
These systems recommend visualizations based on specific goals or aesthetic properties. 
For example, Voyager suggests visualizations emphasizing visual appeal, while SeeDB highlights differences between datasets. LUX~\cite{DBLP:journals/pvldb/LeeTABCKMSYHP21} enables visualization recommendations for dataframes in computational notebooks while factoring in user intent.
 Zenvisage~\cite{DBLP:journals/pvldb/SiddiquiKLKP16} generalizes these systems by enabling the detection of desired visual patterns across large datasets. 
Additionally, systems like Profiler~\cite{DBLP:conf/avi/KandelPPHH12} and Scorpion~\cite{DBLP:journals/pvldb/0002M13} focus on identifying specific patterns, such as outliers, while VizDeck~\cite{DBLP:conf/sigmod/KeyHPA12} presents a dashboard of potential 2D visualizations for a dataset.  
Transactional Panorama~\cite{DBLP:journals/pvldb/TangFGP23} studies the problem of refreshing visualization results.
ML-based systems propose a different approach by learning from data instead of applying predefined rules. 
Draco-Learn~\cite{DBLP:journals/tvcg/MoritzWNLSHH19} optimizes visualization recommendations by learning weights for trade-offs between design constraints. 
VizML~\cite{DBLP:conf/chi/HuBLKH19} focuses on predicting design choices for visualizations, offering better interpretability and ease of integration.

The above studies demonstrate various strategies for enhancing the visualization creation process, either by encoding best practices or learning directly from data with specific contributions to visualization generation and recommendation. 
However, all these methods operate on a \emph{single} database represented by a snowflake schema, whereas \name operates over a collection of related tables from data lakes w.r.t. the given query table.

\section{Conclusion}\label{sec-conc}

In this paper, we studied the new research problem of visualization recommendation for data discovery over data lakes.
We first came up with a formal definition of this problem by addressing the issues of (i) defining how to build visualization over multiple result tables in the output of data discovery; and (ii) developing the concept of series to illustrate the relatedness in data discovery result; and (iii) handling multiple data formats including categorical, numerical and textual columns.
Then, we proposed an end-to-end framework \name as the solution to this problem, which features the technical contributions of a data-driven approach to create the potential series of the visualization as well as progressive pruning strategies to remove unpromising visualization plans.
Experimental results on three public benchmarking tasks for data discovery demonstrated the efficiency of our proposed framework.
Furthermore, our user study shows that the visualization produced by our framework could help solve real data analysis problems in a much shorter time, paving the way toward rapid visual analysis over data lakes.

\bibliographystyle{ACM-Reference-Format}
\bibliography{ref/visual,ref/datalake,ref/other}

\end{document}